\documentclass[a4paper,11pt]{article}

\usepackage[utf8]{inputenc}
\usepackage[T1]{fontenc}
\usepackage[english]{babel}
\usepackage[a4paper,margin=1in]{geometry}
\usepackage{amsmath,amssymb,amsthm,mathtools}
\usepackage{bbm}
\usepackage{graphicx}
\usepackage{booktabs}
\usepackage{threeparttable}
\usepackage{multirow}
\usepackage{makecell}
\usepackage{enumitem}
\usepackage{float}
\usepackage{tikz}
\usepackage{caption}
\usepackage[ruled,vlined,linesnumbered]{algorithm2e}
\usepackage[round,authoryear]{natbib}
\usepackage{orcidlink}
\usepackage{xcolor}
\usepackage{authblk}
\def\bSig\mathbf{\Sigma}

\newcommand{\logit}{\operatorname{logit}}
\newcommand{\expit}{\operatorname{expit}}
\newcommand{\E}{\mathbb E}
\renewcommand{\P}{\mathbb P}
\newcommand{\Var}{\operatorname{Var}}
\newcommand{\bernoulli}{\operatorname{Bernoulli}}
\newcommand{\Cov}{\operatorname{Cov}}
\newcommand{\R}{\mathbb R}
\newcommand{\RD}{\mathrm{RD}}
\newcommand{\RR}{\mathrm{RR}}

\newcommand{\indep}{\perp\!\!\!\perp}
\newcommand{\weil}[1]{\text{\normalfont (#1)}}
\newcommand{\weeil}[2]{\begin{aligned}[t]&\text{\normalfont (#1}\\&\text{\normalfont \phantom{(}#2)}\end{aligned}}

\newtheorem{prop}{Proposition}
\newtheorem{coro}{Corollary}

\title{Quantifying and Calibrating Precision Gain From Covariate Adjustment in Clinical Trials with Binary Endpoints}
\author{Jannik Hammerschmied\orcidlink{0009-0003-0368-5279}}
\author{Christian Stock\orcidlink{0000-0002-3493-3234}}
\author{Martin Schnuerch\orcidlink{0000-0001-6531-2265}\thanks{Correspondence: \href{martin.schnuerch@boehringer-ingelheim.com}{martin.schnuerch@boehringer-ingelheim.com}}}

\affil{Global Biostatistics and Data Sciences, Boehringer Ingelheim Pharma GmbH \& Co. KG, Germany}

\date{}

\begin{document}

\raggedbottom

\maketitle

\begin{abstract}
\noindent Covariate adjustment is widely used in randomized clinical trials to improve estimation and inference of treatment effects. 
For continuous endpoints, the reasoning is straightforward: prognostic baseline covariates explain outcome variability and thereby increase precision. 
For binary endpoints, however, the situation is more complex because conditional and marginal effects generally differ in nonlinear outcome models such as logistic regression. 
We develop a theoretical framework for quantifying the maximum attainable first-order precision gain from covariate adjustment when estimating marginal treatment effects for binary endpoints. 
We derive a closed-form identity showing that precision-gain potential is governed by an estimand- and allocation-specific combination of the treatment-specific risk functions. 
Building on this identity, we introduce marginal prognostic strength as a population-level measure of prognostic information and characterize when it determines precision-gain potential and when the two quantities decouple. 
Specifically, substantial prognostic information within both treatment arms need not translate into improved precision for the marginal treatment effect when covariates are predictive on the target effect scale. 
We further develop a calibration algorithm for methodological simulation studies that preserves prespecified conditional outcome structures and marginal risks while allowing data-generating mechanisms to be calibrated by marginal prognostic strength and evaluated in terms of estimand-specific precision-gain potential. 
Together, these results provide a theoretical and practical framework for understanding, quantifying, and calibrating potential precision gains from covariate adjustment for binary endpoints.
\end{abstract}


\section{Introduction} 
Covariate adjustment is widely used to improve the precision of treatment-effect estimation in clinical trials. 
Improved precision leads to more reliable inference and may reduce the required trial size and duration, benefiting trial sponsors and society alike.
Consequently, regulatory guidance explicitly encourages adjustment for prognostic baseline covariates \citep{EMA2015,FDA2023}.
The underlying intuition is simple: If baseline covariates are prognostic,  accounting for them should remove explainable outcome variation and thereby yield more precise estimates.
Guidance from the US Food and Drug Administration, for instance, states that adjustment ``will generally reduce the variability of estimation of treatment effects and thus lead to narrower confidence intervals and more powerful hypothesis testing'' \citep[p.~3]{FDA2023}.
A substantial body of methodological and empirical literature supports this intuition \citep{BenkeserEtAl2021,KahanEtAl2014,ZhangEtAl2008}.

While covariate adjustment is well understood in the linear regression setting \citep{Lin2013}, its role is more delicate for binary endpoints.
Standard models for binary data yield parameters that reflect conditional treatment effects. 
The treatment coefficient in a logistic regression model, for example, is a conditional log odds ratio.  Although conditional effects are, in principle, most relevant from an individual-patient perspective, they depend on the specified outcome model and may be difficult to estimate consistently in a clinical trial. 
Marginal estimands, in contrast, quantify the population-average treatment effect and are also targeted by assumption-lean, unadjusted analyses.
For these reasons, primary inference in confirmatory trials is typically directed at marginal rather than conditional effects \citep{Lancker2024}.

Whereas marginal and conditional estimands coincide in linear models without treatment-by-covariate interactions, this correspondence no longer holds in generalized linear models with nonlinear link functions, such as logistic regression \citep{DanielEtAl2021}. 
This divergence has motivated the increasing use of standardized, model-assisted estimators such as \(g\)-computation, augmented inverse probability weighting, and targeted maximum likelihood estimation \citep{BannickEtAl2025,MooreVanderLaan2009,Tsiatis2008,Lancker2024,YeEtAl2023}.
In these approaches, regression or conditional-mean working models serve to construct estimators of marginal treatment effects.
Despite the different setting, the guiding intuition remains the same as for linear models: As long as the baseline covariates are prognostic, adjusting for them should increase the efficiency of marginal treatment-effect estimators.

The intuition is valid for estimation of arm-specific risks themselves. 
Covariates that improve the prediction of a binary potential outcome $Y^a$, where $a=1$ and $a=0$ denote the treatment and control arms, respectively, also increase the precision available for estimating the corresponding marginal risk $\mu_a=\E(Y^a)$.
It is therefore natural to expect that improving the arm-specific outcome models will likewise improve precision for marginal treatment-effect estimators constructed from \(\mu_1\) and \(\mu_0\). 
How strongly such improvements in prediction translate into increased precision for treatment-effect estimation, however, remains less well understood.
More importantly, as we show herein, improved arm-specific prediction does not necessarily yield a more precise treatment-effect estimator.

Related work has approached the question of precision gain from a different perspective. 
\citet{Li2023} considered prospective estimation of the efficiency gain from covariate adjustment in future clinical trials using external data. 
These authors defined efficiency parameters under a sharp null, where the treatment-arm-specific conditional outcome functions coincide, and used these to approximate efficiency under local alternatives. 
Our focus is complementary and particularly motivated by methodological simulation studies, where the data-generating mechanism is specified by design.
This enables us to characterize first-order precision-gain potential directly under a given binary-outcome data-generating mechanism, allowing the arm-specific risk functions to differ and treatment effects to vary across covariate values. 
This distinction is important because arm-specific prognostic information may then combine differently across marginal estimands and need not translate directly into precision gain.

The purpose of this article is to characterize the mechanism that determines the precision gain from covariate adjustment for binary endpoints. 
We first derive an identity for the maximal first-order variance reduction attainable through adjustment for a given covariate set when the target estimand is a differentiable function of the treatment-specific marginal risks. 
This identity provides a closed-form expression for the precision-gain potential and shows how it depends jointly on the treatment-arm-specific outcome-covariate relationships, the treatment allocation ratio, and the target estimand. 
We then examine its relationship with marginal prognostic strength, a population-level measure of the information that baseline covariates provide about the outcome, and show why prognostic strength alone does not determine the potential precision gain. 
Building on these results, we develop an algorithm for methodological simulation studies that calibrates the data-generating mechanism to a specified level of marginal prognostic strength while preserving the arm-specific coefficient structure and marginal risks, and quantifies the resulting precision-gain potential for the marginal estimand and treatment allocation of interest.
Finally, we illustrate the use of the algorithm in the design of a simulation study.

\section{Motivating Example}
\label{sec:example}

We begin with an example that challenges the usual prognostic-covariate intuition under conditions favorable for adjustment. 
Consider a two-arm randomized controlled trial comparing an investigational treatment ($A = 1$) with control ($A = 0$) in the presence of a binary baseline covariate $X$. 
Let 
$$ X\sim \bernoulli(0.85), \qquad A\sim \bernoulli(0.50),$$
with $A \indep (X, Y^0, Y^1)$ due to randomization. Conditional on treatment and covariate, the binary endpoint follows 
$$ Y\mid A,X \sim \bernoulli\left\{p(A,X)\right\},$$ 
with $\logit\left\{p(A,X)\right\} = 0.25-0.5A+3.3X-6.6AX$.
Treatment lowers the outcome risk compared with control at both covariate levels, although the magnitude of the treatment effect differs substantially between them. By conventional coefficient-based descriptions, \(X\) is strongly prognostic in both treatment arms: its conditional log-odds contrast is \(3.3\) in the control arm and \(3.3-6.6=-3.3\) in the treatment arm, corresponding to conditional odds ratios of approximately \(27\) and \(1/27\), respectively. One might therefore expect adjustment for \(X\) to be especially useful for estimating the marginal treatment effect. 

We compare an unadjusted estimator with a logistic \(g\)-computation estimator based on the correctly specified, saturated working model 
$$ \logit\left\{\E(Y\mid A,X)\right\} = \beta_0+\beta_AA+\beta_XX+\beta_{AX}AX, $$
with marginal risks estimated by standardization, 
$$\hat\mu_a = \frac{1}{n}\sum_{i=1}^n \logit^{-1} \left( \hat\beta_0+\hat\beta_Aa+\hat\beta_X X_i+\hat\beta_{AX}aX_i \right), \qquad a\in\{0,1\}.$$ 
Thus, the adjusted analysis is considered under favorable conditions: the outcome model is correctly specified and saturated, and the marginal estimand is targeted through standardization. Under these conditions, logistic \(g\)-computation is consistent for the marginal risks \(\mu_0,\mu_1\) and asymptotically efficient, attaining the nonparametric efficiency bound for the marginal risks and any continuously differentiable function thereof \citep{RosenblumVanderLaan2010}. 

We first consider the marginal risk difference, $\psi_{RD}=\mu_1-\mu_0$.
To assess the precision gain from adjustment, we simulate \(100{,}000\) trials with \(n=500\) participants each and compute both estimators in every trial. 
We estimate their Monte Carlo variances and the relative variance reduction of the adjusted versus the unadjusted estimator, 
$$ 1- \frac{ \widehat{\Var}_{\mathrm{MC}}(\hat\psi_{\mathrm{adj}}) }{ \widehat{\Var}_{\mathrm{MC}}(\hat\psi_{\mathrm{unadj}}) }.$$
The results are shown in the first column of Table~\ref{tab:intro}. 

\begin{table}[ht] 
\centering 
\begin{threeparttable} 
\caption{ Monte Carlo comparison of unadjusted and adjusted marginal treatment-effect estimators } 
\label{tab:intro} 
\begin{tabular}{lrr} 
\toprule 
& Risk Difference & Risk Ratio \\ 
\midrule 
Unadjusted estimator & $6.55\cdot10^{-4}$ & $4.01 \cdot 10^{-4}$ \\ 
Logistic $g$-computation & $6.58\cdot 10^{-4}$ & $3.60\cdot 10^{-4}$ \\ 
\midrule
Relative variance reduction & $-0.5\%$ & $10.1\%$ \\ 
\bottomrule 
\end{tabular} 
\begin{tablenotes} 
\footnotesize \item \textit{Note.} The logistic g-computation model converged in all simulated trials, and all trials were included in the Monte Carlo summaries.
\end{tablenotes} 
\end{threeparttable} 
\end{table} 

Despite the strong arm-specific associations between \(X\) and the outcome and the favorable conditions for adjustment, logistic \(g\)-computation provides essentially no precision gain for the marginal risk difference in this example. 
The picture changes when the target estimand is the marginal risk ratio, $\psi_{RR}=\mu_1/\mu_0$.
Under the same data-generating mechanism and using the same adjusted analysis, the Monte Carlo variance is reduced by \(10.1\%\) relative to the unadjusted estimator (Table~\ref{tab:intro}, second column). 
Thus, the same covariate that provides essentially no precision gain for the risk difference yields a substantial gain for the risk ratio. 

To summarize, the example shows that the benefit of covariate adjustment is not a property of the covariate alone.
Under the same data-generating mechanism and under the same, correctly specified adjusted analysis, a covariate can provide essentially no precision gain for one estimand, yet a meaningful gain for another estimand. 
The difference cannot be attributed to weak arm-specific associations or model misspecification, illustrating that the precision gain from covariate adjustment is determined by factors beyond the apparent prognostic strength of the covariate.

\section{Potential Precision Gain From Covariate Adjustment}
\label{sec:identity}
The preceding example demonstrates that a single, informal appeal to a covariate being
``prognostic'' cannot account for when adjustment actually improves treatment-effect estimation precision.
The fundamental question remains: How much first-order precision can, in principle, be gained by adjusting for a given set of baseline covariates when estimating a specified marginal treatment effect?
We address this question by defining \emph{precision-gain potential} as the maximal first-order variance reduction attainable through covariate adjustment and deriving a general identity that characterizes this quantity.

\subsection{The Precision-Gain Identity for Differentiable Marginal Estimands}

Throughout this article, we consider real-valued marginal estimands, that is, targeted treatment effects in the study population. 
Specifically, 
$$\psi=f(\mu_1,\mu_0), \qquad \mu_a=\E(Y^a), \qquad a\in\{0,1\},$$
where $f$ is differentiable at $(\mu_1,\mu_0)$.
This class of estimands includes common marginal treatment effects for binary endpoints, such as the risk difference, risk ratio, and odds ratio.

We consider i.i.d.\ observations \(\{(X_i,A_i,Y_i)\}_{i=1}^n\) from a randomized trial, where $X$ denotes the vector of baseline covariates used for adjustment.  We assume consistency, \(Y=AY^1+(1-A)Y^0\), and randomization, $A\indep(X,Y^0,Y^1)$, with known randomization probability $\pi = \P(A =1)\in(0,1)$. The corresponding arm-specific conditional risk functions are
$$p_a(X)=\E(Y^a\mid X).$$

To quantify the precision gain available from adjustment for $X$, let $\hat \mu_a$ denote the unadjusted, empirical risk in treatment arm $a$, and define the unadjusted estimator $\hat \psi_\text{unadj} = f(\hat \mu_1, \hat \mu_0)$. Its first-order asymptotic variance constant is denoted by \(\nu_{\mathrm{unadj}}(\psi)\), and the smallest first-order asymptotic variance constant attainable by a regular, asymptotically linear estimator using \((X,A,Y)\) is denoted by \(\nu^\star_{\mathrm{adj}}(\psi;X)\). 
We define the \emph{absolute precision-gain potential} as 
$$\mathcal{G}^\mathrm{abs}_{\pi,\psi}(X) = \nu_{\mathrm{unadj}}(\psi) - \nu^\star_{\mathrm{adj}}(\psi;X),$$ 
with corresponding \emph{relative precision-gain potential} 
$$\mathcal{G}^\mathrm{rel}_{\pi,\psi}(X) = 1- \frac{ \nu^\star_{\mathrm{adj}}(\psi;X) }{ \nu_{\mathrm{unadj}}(\psi) }.$$
Precision-gain potential thus quantifies what adjustment for $X$ can ultimately achieve in terms of precision: 
It tells us how much asymptotic variance can at most be removed in the given setting. 
The following proposition provides a closed-form expression for this limit and reveals the factors that govern precision gain.

\begin{prop}[Precision-Gain Identity]
\label{prop:gain-identity}
Let $f:\mathbb R^2\to\mathbb R$ be differentiable at $(\mu_1,\mu_0)$. Define $(f_1,f_0)^\top=\nabla f(\mu_1,\mu_0)$, and suppose \((f_1,f_0)\neq(0,0)\). For \(\psi=f(\mu_1,\mu_0)\), the absolute and relative precision-gain potential are given, respectively, by
\[
\mathcal{G}^\mathrm{abs}_{\pi,\psi}(X)
=
\Var\left(
\sqrt{\frac{1-\pi}{\pi}}\,f_1p_1(X)
-
\sqrt{\frac{\pi}{1-\pi}}\,f_0p_0(X)
\right)
\]
and
\[
\mathcal{G}^\mathrm{rel}_{\pi,\psi}(X)=\frac{\Var\left(
\sqrt{\frac{1-\pi}{\pi}}\,f_1p_1(X)
-
\sqrt{\frac{\pi}{1-\pi}}\,f_0p_0(X)
\right)}{f_1^2\frac{\Var(Y^1)}{\pi}+f_0^2\frac{\Var(Y^0)}{1-\pi}}.
\]
\end{prop}
Proofs of Proposition \ref{prop:gain-identity} and all subsequent theoretical results are provided in the Appendix.
The identity follows from standard semiparametric efficiency theory for randomized trials \citep{Tsiatis2008}.
Our representation makes explicit how the arm-specific conditional outcome functions combine based on the target estimand and treatment allocation. 
Although Proposition~1 may apply more generally to other outcomes by replacing $p_a(X)$ with $\E(Y^a\mid X)$, we retain the binary-outcome formulation throughout because the subsequent development is specific to binary endpoints.

The identity shows that precision-gain potential is determined by the variability of an estimand- and allocation-specific combination of the two arm-specific risk functions. 
Thus, how much precision can be gained depends not only on how the covariates relate to the outcome within each treatment arm, but also on the target estimand and the randomization probability.

Interestingly, the precision-gain identity constitutes a generalization of an earlier result of \citet[Section~6]{MooreVanderLaan2009} for the relative efficiency of targeted maximum likelihood estimation. For the marginal risk difference and under equal randomization, they expressed the absolute variance reduction as
$$4\left(\E\left[\left\{Y-\E(Y)\right\}^2\right]-\E\left[\left\{Y-\E(Y\mid X)\right\}^2\right]\right),$$
which equals $4\Var\left\{\E(Y\mid X)\right\}$ by the law of total variance. This result can be recovered directly from Proposition~\ref{prop:gain-identity}: For the risk difference, $f_1 = 1$ and $f_0 = -1$, and under equal randomization, $\pi = 1/2$, so that $\E(Y\mid X)=\left\{p_1(X)+p_0(X)\right\} / 2$. Hence,
$$\mathcal{G}^{\mathrm{abs}}_{1/2,\mathrm{RD}}(X)=\Var\left\{p_1(X)+p_0(X)\right\}=4\Var\{\E(Y\mid X)\}.$$
The result of \citet{MooreVanderLaan2009} is thus a special case of the precision-gain identity derived herein.

\section{From Prognostic Strength to Precision Gain}
\label{sec:notions}
The precision-gain identity establishes how much precision for estimation of a marginal estimand can be gained through covariate adjustment and what determines this limit. 
It also clarifies why informal descriptions of covariates as ``prognostic'' are insufficient for this purpose: precision-gain potential depends on how the two arm-specific risk functions combine for the target estimand and treatment allocation. 
We therefore revisit the usual prognostic-covariate intuition by introducing an arm-specific, population-level measure of prognostic information. 
We then relate this quantity to conventional coefficient-based descriptions of prognostic associations and characterize when prognostic information determines precision-gain potential, and when it does not.

\subsection{Marginal prognostic strength}

To isolate the prognostic information carried by $X$ within a treatment arm, consider estimation of a single marginal risk $\mu_a$. 
Applying Proposition~\ref{prop:gain-identity} with $\psi=\mu_a$ yields 
$$\mathcal{G}^{\mathrm{abs}}_{\pi,\mu_a}(X) = \frac{\P(A=1-a)}{\P(A=a)} \Var\left\{p_a(X)\right\}.$$ 
The first factor is design-specific and determined by the treatment allocation, whereas $\Var\left\{p_a(X)\right\}$ captures how informative the covariates are for the outcome in treatment arm $a$, accounting for both the arm-specific risk function and the population distribution of the covariates. 
We therefore define 
$$\mathcal{S}^{\mathrm{abs}}_a(X) = \Var\left\{p_a(X)\right\} $$ 
as the \emph{absolute marginal prognostic strength} of $X$ in treatment arm $a$.
Its relative counterpart is 
$$\mathcal{S}^{\mathrm{rel}}_a(X) = \frac{\Var\{p_a(X)\}}{\Var(Y^a)} = \frac{\Var\{p_a(X)\}}{\mu_a(1-\mu_a)}.$$ 

Marginal prognostic strength has a direct prediction-based interpretation: 
Under squared-error loss, by the $L^2$-projection property of conditional expectation, the maximal reduction in population prediction error achievable by using $X$ rather than the marginal risk alone is
$$\E\left\{(Y^a-\mu_a)^2\right\} - \E\left[\left\{Y^a-p_a(X)\right\}^2\right] = \Var\left\{p_a(X)\right\} = \mathcal{S}^{\mathrm{abs}}_a(X).$$
Accordingly, $\mathcal{S}^{\mathrm{rel}}_a(X)$ is the proportion of marginal outcome variance explained by the arm-specific conditional risk function. 
Marginal prognostic strength therefore provides a population-level measure of the prognostic information carried by $X$, accounting for both the arm-specific risk function and the distribution of the covariates.

Of note, the prediction-based interpretation of relative marginal prognostic strength reveals a direct correspondence with the \emph{coefficient of discrimination} proposed independently by \citet{Tjur2009}. 
Specifically,
$$\mathcal S_a^{\mathrm{rel}}(X) = \E\{p_a(X)\mid Y^a=1\} - \E\{p_a(X)\mid Y^a=0\}.$$
Thus, the population quantity that emerges here from the theory of covariate-adjustment efficiency coincides with the population analogue of Tjur's coefficient of discrimination. 
Importantly, our formulation of marginal prognostic strength is not restricted to logistic regression and embeds this measure in a framework that makes its relationship to precision-gain potential for marginal treatment-effect estimation explicit.

\subsection{Why Outcome-Model Coefficients Do Not Determine Marginal Prognostic Strength} 

A common, informal way to describe prognostic strength is through parameters of a conditional outcome model. 
For binary outcomes, however, coefficient magnitude alone does not determine marginal prognostic strength. 
Consider a binary covariate $X \sim \bernoulli(q)$ and an arm-specific logistic model 
$$\logit\left\{p_a(X)\right\} = \alpha_a+\beta_aX.$$ 
Then,
$$\mathcal{S}^{\mathrm{abs}}_a(X) = q(1-q)\left\{p_a(1)-p_a(0)\right\}^2.$$ 
Hence, marginal prognostic strength depends not only on the conditional log-odds contrast $\beta_a$, but also on the baseline risk through $\logit^{-1}\left\{\alpha_a\right\} = p_a(0)$ and on the covariate distribution through $q$.
Consequently, the magnitude of a conditional model coefficient does not by itself characterize how much prognostic information a covariate carries at the population level.

This distinction resolves one ambiguity in informal descriptions of covariates as ``strongly prognostic.'' 
Coefficient-based associations describe prognostic relevance on the scale of a conditional outcome model, whereas marginal prognostic strength quantifies the prognostic information carried by the covariates in the study population. 
The two should not be used interchangeably.

\subsection{When Does Marginal Prognostic Strength Determine Precision Gain?}

Marginal prognostic strength quantifies the prognostic information carried by the covariates separately within each treatment arm. 
Precision-gain potential for a marginal treatment effect, in contrast, depends on how the two arm-specific risk functions combine in the precision-gain identity. 
A direct relationship between the two quantities emerges when the covariates are non-predictive on the target effect scale. 

In clinical-trial terminology, a baseline covariate is described as predictive if it is associated with the treatment effect. 
Because the treatment effect depends on the chosen effect scale, predictiveness is necessarily scale-specific. 
For an effect scale $f$, we say that $X$ is not predictive on the $f$-scale if $f(p_1(X),p_0(X))$ is constant.
Thus, for the risk difference, non-predictiveness means that $p_1(X)-p_0(X)$ is constant, whereas for the risk ratio it means that ${p_1(X)}/{p_0(X)}$ is constant. 
In either case, the two arm-specific risk functions are affinely related, so that their contributions to the precision-gain identity cannot vary independently. 
In these cases, marginal prognostic strength directly determines precision-gain potential.
\begin{coro}[Precision-Gain Potential Under Non-Predictiveness]
\label{coro:nonpredictive}
Suppose $X$ is not predictive on the target effect scale. Then,
\begin{enumerate}
\item for the marginal risk difference,
\[
\mathcal{G}^\mathrm{abs}_{\pi,\mathrm{RD}}(X)
=
\frac{1}{\pi(1-\pi)}\,\mathcal{S}^\mathrm{abs}_{0}(X)
=
\frac{1}{\pi(1-\pi)}\,\mathcal{S}^\mathrm{abs}_{1}(X),
\]
\item for the marginal risk ratio, with $\rho=\mu_1/\mu_0$ and $\mu_0>0$,
\[
\mathcal{G}^\mathrm{abs}_{\pi,\mathrm{RR}}(X)
=
\frac{\rho^2}{\mu_0^2\,\pi(1-\pi)}\,\mathcal{S}^\mathrm{abs}_{0}(X)
=
\frac{1}{\mu_0^2\,\pi(1-\pi)}\,\mathcal{S}^\mathrm{abs}_{1}(X),
\]
\item for both estimands, provided $\mathcal{S}^\mathrm{rel}_{1}(X),\mathcal{S}^\mathrm{rel}_{0}(X)>0$,
\[
\mathcal{G}^\mathrm{rel}_{\pi,f}(X)
=
\left(\frac{1-\pi}{\mathcal{S}^\mathrm{rel}_{1}(X)}+\frac{\pi}{\mathcal{S}^\mathrm{rel}_{0}(X)}\right)^{-1},
\qquad f\in\{\mathrm{RD},\mathrm{RR}\}.
\]
\end{enumerate}
\end{coro}
Corollary~\ref{coro:nonpredictive} identifies an important setting in which the usual prognostic-covariate intuition is exact. 
When the covariates are non-predictive on the risk-difference or risk-ratio scale, greater marginal prognostic strength implies greater precision-gain potential for the corresponding marginal treatment effect. 
In these settings, arm-specific prognostic information therefore translates directly into the potential for precision gain, with relative precision-gain potential being the reverse-allocation-weighted harmonic mean of the arm-specific relative marginal prognostic strengths.

For the marginal odds ratio, in contrast, non-predictiveness induces a nonlinear relationship between the arm-specific risk functions, and marginal prognostic strength therefore does not generally determine precision-gain potential. 
Concrete counterexamples are provided in the Appendix (Section B.2).

A similar disconnect can arise for the risk difference and risk ratio when the covariates are predictive on the target effect scale.
Variation in the two arm-specific risk functions may then offset one another. 
Consequently, substantial marginal prognostic strength in both treatment arms can coexist with little or no precision gain for the treatment effect.

This mechanism is best illustrated with a simple example. 
Let $X \sim \bernoulli(1/2)$ and suppose 
$$Y^1=X, \qquad Y^0=1-X.$$ 
The covariate then perfectly predicts both potential outcomes, such that 
$$\mathcal{S}^{\mathrm{rel}}_1(X) = \mathcal{S}^{\mathrm{rel}}_0(X) = 1.$$ 
However, for the marginal risk difference under equal randomization, 
$$\mathcal{G}^{\mathrm{abs}}_{1/2,\mathrm{RD}}(X) = \Var\left\{p_1(X)+p_0(X)\right\} = \Var\{X+(1-X)\}=0.$$ 
Thus, even perfect arm-specific prognostic information need not yield any precision gain for a marginal treatment effect if the arm-specific risk functions vary in opposite directions.
Table \ref{tab:relationships} provides an overview of the relationship between marginal prognostic strength and precision-gain potential across different scenarios. 

\begin{table}[ht]
\centering

\caption{Relationship between marginal prognostic strength and precision-gain potential}
\label{tab:relationships}

\begin{threeparttable}
\begin{tabular}{llc}
\toprule
Covariate status
& Estimand
& MPS $\dashrightarrow$ PGP \\
\midrule

\multirow{3}{*}{\shortstack[l]{Not predictive on target scale}}
& RD & \checkmark \\
& RR & \checkmark \\
& OR & $\times$ \\
\midrule

\multirow{3}{*}{\shortstack[l]{Predictive on target scale}}
& RD & $\times$ \\
& RR & $\times$ \\
& OR & $\times$ \\
\bottomrule
\end{tabular}

\begin{tablenotes}
\footnotesize
\item \textit{Note.} MPS = marginal prognostic strength;
PGP = precision-gain potential; RD = marginal risk difference; RR = marginal risk ratio; OR = marginal odds ratio. The notation \(A\dashrightarrow B\) asks whether \(A\) uniquely determines \(B\).
\end{tablenotes}
\end{threeparttable}

\end{table}

The motivating example in Section~\ref{sec:example} illustrates the same phenomenon in a less extreme setting. 
Here, $X$ has a relative marginal prognostic strength of approximately $26.3\%$ in both treatment arms. 
For the marginal risk difference, however, the two arm-specific risk functions vary in opposite directions and offset in the precision-relevant combination, resulting in zero precision-gain potential. 
For the marginal risk ratio, this cancellation does not occur, yielding a precision-gain potential of approximately $10.6\%$. 
These asymptotic results are closely reflected by the finite-sample Monte Carlo results in Table~\ref{tab:intro}, with relative variance reductions of $-0.5\%$ and $10.1\%$, respectively.
Thus, even after prognostic information is quantified at the population level, whether and to what extent it translates into precision gain depends on how the arm-specific risk functions combine for the given target estimand and treatment allocation.

\section{Structure-Preserving Calibration of Data-Generating Mechanisms} 
\label{sec:calibration} 

The preceding results have direct implications for the design of methodological simulation studies evaluating covariate-adjusted methods. 
Specifying large outcome-model coefficients does not ensure substantial marginal prognostic strength and, more importantly, does not necessarily increase the precision gain attainable for the marginal treatment effect under study. 
Simulation scenarios defined primarily through coefficient magnitude may therefore have substantially different efficiency properties than intended. 

To address this problem, we formally develop a structure-preserving calibration algorithm for constructing data-generating mechanisms with controlled marginal properties and establish the theoretical properties required for a well-defined calibration procedure.
The user specifies the conditional outcome-model coefficients and desired marginal risks. 
The procedure then generates a structure-preserving calibration path by jointly scaling the arm-specific coefficients while recalibrating the intercepts to maintain the marginal risks. 
Along this path, each attainable control-arm marginal prognostic strength identifies a unique scaled data-generating mechanism with corresponding treatment-arm prognostic strength and estimand-specific precision-gain potential. 
This enables simulation scenarios to be constructed with controlled and interpretable marginal prognostic strength and estimand-specific precision-gain potential, rather than relying on coefficient magnitude as a proxy.

The formal development below establishes the key properties underlying this calibration procedure. 
We show that marginal prognostic strength varies strictly monotonically along the structure-preserving path and therefore uniquely indexes the corresponding data-generating mechanisms, thereby reparameterizing the unbounded coefficient-scaling space by a bounded and interpretable calibration coordinate. 
We further characterize the upper limit of this calibration coordinate. 
These results provide the theoretical basis for Algorithm~\ref{alg:structure-preserving-calibration}, which translates the developed framework into a practical calibration procedure.

\subsection{Calibration Framework} 

Let $X$ follow a prespecified target covariate distribution, and let 
$$h(X) = \left\{h_1(X),\ldots,h_d(X)\right\}^{\mathsf T}$$ 
denote a user-specified vector of covariate functions, which may include the original baseline covariates as well as nonlinear transformations and interactions.
The conditional outcome structure is specified through 
$$\logit\left\{\E(Y\mid A,X)\right\} = \alpha_0+\alpha_A A + v_{\mathrm{hom}}^{\mathsf T}h(X) + A v_{\mathrm{het}}^{\mathsf T}h(X),$$
where $v_{\mathrm{hom}}\in\mathbb R^d$ denotes the coefficient component shared by both treatment arms and $v_{\mathrm{het}}\in\mathbb R^d$ the additional treatment-interaction component. 
The coefficient directions in the control and treatment arms are therefore 
$v_{\mathrm{hom}}$ and $v_{\mathrm{hom}}+v_{\mathrm{het}}$, respectively. 
Their signs, zero patterns, and relative magnitudes specify the qualitative conditional association and treatment-heterogeneity structure. 

In addition, the user specifies the desired arm-specific marginal risks $0<\mu_0,\mu_1<1$. 
The intercepts are not retained from the initial outcome model but are calibrated to preserve these marginal risks throughout the procedure. 

\subsection{A Structure-Preserving Calibration Path} 

The user-specified coefficients define the qualitative conditional structure but do not directly reveal its marginal properties and are impractical for manually controlling them.
We therefore introduce a common multiplier $\lambda \geq 0$ and define 
$$p_{0,\lambda}(X) = \expit\left\{ \alpha_0(\lambda) + \lambda v_{\mathrm{hom}}^{\mathsf T}h(X) \right\},$$ 
and 
$$p_{1,\lambda}(X) = \expit\left\{ \alpha_1(\lambda) + \lambda (v_{\mathrm{hom}}+v_{\mathrm{het}})^{\mathsf T}h(X) \right\}.$$ 
For each $\lambda$, the arm-specific intercepts are calibrated such that 
$$\E\left\{p_{a,\lambda}(X)\right\} = \mu_a, \qquad a\in\{0,1\}.$$
Varying $\lambda$ therefore changes the overall magnitude of the conditional associations while preserving their signs, zero patterns, relative magnitudes, and treatment-heterogeneity structure, as well as the marginal risks. 
This defines the one-dimensional structure-preserving path 
$$\mathcal P = \left\{ (p_{0,\lambda},p_{1,\lambda}) : \lambda\in[0,\infty) \right\}.$$ 
Calibration now amounts to selecting a data-generating mechanism along $\mathcal P$. 
Because $\lambda$ has no direct marginal interpretation and no finite upper bound, it is not a convenient calibration coordinate. 
We therefore reparameterize the path by relative marginal prognostic strength, which provides an interpretable and bounded marginal scale.

\subsection{Marginal Prognostic Strength as the Calibration Coordinate} 

For the relative marginal prognostic strength to serve as a calibration coordinate, each possible value must identify a unique point on the structure-preserving path. 
In the following, we establish this property.

Define the arm-specific directional scores 
$$\eta_0(X) = v_{\mathrm{hom}}^{\mathsf T}h(X), \qquad \eta_1(X) = (v_{\mathrm{hom}}+v_{\mathrm{het}})^{\mathsf T}h(X).$$
Along the structure-preserving path, the relative marginal prognostic strength in arm $a$ is
$$\mathcal S_{a,\lambda}^{\mathrm{rel}}(X) = \frac{ \Var\{p_{a,\lambda}(X)\} }{ \mu_a(1-\mu_a) }, \qquad a\in\{0,1\}.$$

\begin{prop}[Monotonicity of Marginal Prognostic Strength] 
\label{prop:mps-monotonicity} 
Let $a \in \{0,1\}$ and suppose that $\eta_a(X)$ is nondegenerate and integrable.
Then the map 
$$\lambda\longmapsto \mathcal S_{a,\lambda}^{\mathrm{rel}}(X)$$ is differentiable on $\mathbb R$ and strictly increasing on $[0,\infty)$. 
\end{prop} 

Proposition~\ref{prop:mps-monotonicity} implies that every attainable control-arm marginal prognostic strength identifies a unique multiplier and hence a unique data-generating mechanism along $\mathcal P$. 
This provides a one-to-one reparameterization of the path from the unbounded coefficient multiplier $\lambda \in [0,\infty)$ to the bounded relative marginal prognostic strength $\mathcal S_{a,\lambda}^{\mathrm{rel}}(X)\in[0,1]$.
The bounded parameterization is essential: whereas direct exploration over $\lambda$ provides no natural finite upper bound, the marginal prognostic-strength scale has a well-defined limiting value. 
To characterize the attainable range of the calibration coordinate, it therefore remains to determine this limit.

\begin{prop}[Limiting Marginal Prognostic Strength] 
\label{prop:mps-limit} 
Let $a\in\{0,1\}$, $0<\mu_a<1$, and let $c\in\mathbb R$ satisfy 
$$\P\{\eta_a(X)>c\} \leq \mu_a \leq \P\{\eta_a(X)\geq c\}.$$
Define 
$$q = \begin{cases} 
\displaystyle \frac{ \mu_a-\P\{\eta_a(X)>c\} }{ \P\{\eta_a(X)=c\} }, & \text{if }\P\{\eta_a(X)=c\}>0, \\[2ex] 
0, & \text{if }\P\{\eta_a(X)=c\}=0. 
\end{cases}$$ 
Then 
$$p_{a,\lambda}(X) \xrightarrow{\mathrm{a.s.}} \mathbf 1_{\{\eta_a(X)>c\}} + q\mathbf 1_{\{\eta_a(X)=c\}} \qquad (\lambda\to\infty),$$ 
and 
$$\lim_{\lambda\to\infty} \Var\{p_{a,\lambda}(X)\} = \P\{\eta_a(X)>c\} + q^2\P\{\eta_a(X)=c\} - \mu_a^2.$$ 
Consequently, 
$$\mathcal S_{a,\infty}^{\mathrm{rel}}(X) := \lim_{\lambda\to\infty}\mathcal S_{a,\lambda}^{\mathrm{rel}}(X) = \frac{ \P\{\eta_a(X)>c\} + q^2\P\{\eta_a(X)=c\} - \mu_a^2 }{ \mu_a(1-\mu_a) }.$$ 
\end{prop} 

The limiting value $\mathcal S_{0,\infty}^{\mathrm{rel}}(X)$ characterizes the upper endpoint of the control-arm calibration range and can be approximated from a Monte Carlo reference sample from the target covariate distribution. 
Together, Propositions~\ref{prop:mps-monotonicity} and~\ref{prop:mps-limit} thus transform the unbounded search over the coefficient multiplier into a bounded calibration problem with a known upper endpoint.
\subsection{Calibration Algorithm} 
\label{sec:calibration-algorithm} 
Because $\mathcal S_{0,\infty}^{\mathrm{rel}}(X)$ is generally approached only in the limit, let $t\in(0,1)$ denote a truncation fraction and construct an equidistant grid 
$$0=s_{1}<s_{2}<\cdots<s_{K}=t \mathcal S_{0,\infty}^{\mathrm{rel}}(X).$$ 
By Propositions~\ref{prop:mps-monotonicity} and~\ref{prop:mps-limit}, every $s_{k}$ identifies a unique multiplier $\lambda_k$ satisfying $\mathcal S_{0,\lambda_k}^{\mathrm{rel}}(X) =s_{k}$.
The arm-specific intercepts are then calibrated to satisfy 
$$\E\{p_{a,\lambda_k}(X)\} = \mu_a, \qquad a\in\{0,1\},$$
yielding a unique pair $(p_{0,\lambda_k},p_{1,\lambda_k})$ with the prespecified qualitative coefficient structure and marginal risks. 

For each particular data-generating mechanism along the calibration path, the treatment-arm marginal prognostic strength and precision-gain potential can be evaluated directly. 
For any differentiable marginal estimand \(\psi=f(\mu_1,\mu_0)\) and treatment-allocation probability \(\pi\in(0,1)\), Proposition~\ref{prop:gain-identity} gives the estimand-specific relative precision-gain potential. 
The resulting calibration path therefore connects three levels of specification: coefficient directions preserve the qualitative conditional outcome structure, marginal prognostic strength provides an interpretable coordinate along the structure-preserving path, and precision-gain potential quantifies the attainable efficiency gain for the estimand and allocation of interest. 
Algorithm~\ref{alg:structure-preserving-calibration} summarizes the complete procedure.

\begin{algorithm}[H] 
\caption{ Structure-preserving calibration of data-generating mechanisms for binary-outcomes} 
\label{alg:structure-preserving-calibration} 
\KwIn{ Reference sample $\{X_i\}_{i=1}^N$; $h$, $v_{\mathrm{hom}}$, $v_{\mathrm{het}}$; $\mu_0,\mu_1$; $\pi$; estimand gradients $(f_1, f_0)$ of interest; $t$; $K$ }
\KwOut{ Calibration path and coefficients of a selected data-generating mechanism}

Evaluate $\eta_{0i} = v_{\mathrm{hom}}^{\mathsf T}h(X_i)$ and $\eta_{1i} = v_{\mathrm{hom}}^{\mathsf T}h(X_i) + v_{\mathrm{het}}^{\mathsf T}h(X_i); $ 

\SetKwFunction{ControlMPS}{ControlMPS} \SetKwFunction{Bisect}{Bisection} \SetKwProg{Fn}{Function}{:}{} \Fn{\ControlMPS{$\lambda$}}{ Obtain $\alpha_0(\lambda)$ by $\alpha_0(\lambda) = \Bisect\!\left( \alpha_{\mathrm{lower}}, \alpha_{\mathrm{upper}}; \frac{1}{N}\sum_{i=1}^N \expit(\alpha+\lambda\eta_{0i})-\mu_0 \right)$\; \Return{ $\displaystyle \frac{1}{N\mu_0(1-\mu_0)} \sum_{i=1}^N (\expit(\alpha_0(\lambda)+\lambda\eta_{0i}) -\mu_0 )^2$ }\; }

Approximate $\mathcal S_{0,\infty}^{\mathrm{rel}}(X)$ from the reference sample using Proposition~\ref{prop:mps-limit}\;
Construct the control-arm marginal prognostic strength grid $0=s_{1}<\cdots<s_{K}=t \mathcal S_{0,\infty}^{\mathrm{rel}}(X);$ 

\For{$k=1,\ldots,K$}{ Obtain $\lambda_k = \Bisect\!\left( \lambda_{\mathrm{lower}}, \lambda_{\mathrm{upper}}; \ControlMPS(\lambda)-s_{k} \right)$\;

Set $\alpha_{0,k}=\alpha_0(\lambda_k)$\; Obtain $\alpha_{1,k}$ by bisection under the treatment-arm marginal-risk constraint\; Evaluate treatment-arm marginal prognostic strength and precision-gain potential for the estimands of interest\; }

Display treatment-arm marginal prognostic strength and precision-gain potential as functions of the attainable control-arm marginal prognostic strength\; Given a user-selected control-arm marginal prognostic strength $s_0^\star$, recalibrate the corresponding multiplier and intercepts and return the selected data-generating mechanism\;
\end{algorithm} 

Although marginal prognostic strength serves as the formal calibration coordinate, the calibration path can also be used to target the magnitude of precision-gain potential. 
By displaying estimand-specific precision-gain potential as a function of control-arm marginal prognostic strength, the path identifies regions corresponding to the desired precision gain.
The user can therefore select a control-arm marginal prognostic strength that yields a suitable precision-gain potential and calibrate the final data-generating mechanism to this value.
Thus, the procedure permits calibration with respect to precision-gain potential without requiring precision-gain potential itself to uniquely index the path.

The finite grid $\{s_1,\ldots,s_K\}$ serves only to construct and display the calibration path.
Once a target control-arm marginal prognostic strength has been selected, the final data-generating mechanism is calibrated directly to that value and is therefore not restricted by the grid resolution.
Marginal prognostic strength provides a unique calibration coordinate because it varies strictly monotonically along the structure-preserving path. 
In contrast, monotonicity does not hold for precision-gain potential. 
Increasing the magnitude of all arm-specific coefficients may cause the attainable precision gain to increase, then decrease and possibly increase again. 
Consequently, the same precision-gain target may occur at multiple points along the path and therefore correspond to different data-generating mechanisms.
A formal characterization of limiting precision-gain potential and an example demonstrating its nonmonotonicity are provided in the Appendix (Section C.3).

To facilitate application of Algorithm~\ref{alg:structure-preserving-calibration} in practice, we provide an implementation in the statistical programming environment R. 
The extensively annotated, ready-to-use script defines functions implementing the individual steps of the algorithm and is accompanied by a worked example illustrating its use in a methodological simulation setting. 
The complete R implementation, including the worked example and comprehensive instructions, is available at \url{https://github.com/Boehringer-Ingelheim/binary-dgm-calibration}.

\section{Illustrative Application} 
\label{sec:illustration} 

We illustrate how the proposed calibration algorithm can be used to construct an informative data-generating mechanism for a methodological simulation study. 
The aim is to preserve a prespecified conditional outcome structure and marginal treatment effect while explicitly controlling the population-level prognostic information and quantifying the resulting estimand-specific potential for precision gain.
The worked example is included with the R implementation described in Section~\ref{sec:calibration-algorithm}, allowing the complete analysis to be reproduced using the accompanying code.

Consider two independent baseline covariates, 
\[ X_1\sim\bernoulli(0.30), \qquad X_2\sim\mathcal N(0,1), \qquad X_1\indep X_2, \] 
with the vector of covariate functions
\[ h(X) = \left( X_1,\, X_2,\, X_1X_2^2 \right)^{\mathsf T}. \] 
We specify the coefficient directions 
\[ v_{\mathrm{hom}} = \left( 0.6,\, 0.4,\, 0.25 \right)^{\mathsf T}, \qquad v_{\mathrm{het}} = \left( 0,\, -1.1,\, 0 \right)^{\mathsf T}, \] 
and marginal risks 
\[ \mu_0=0.55, \qquad \mu_1=0.15. \]
Thus, the conditional outcome structure includes homogeneous associations for \(X_1\), \(X_2\), and \(X_1X_2^2\), together with treatment-effect heterogeneity on the log-odds ratio scale through \(X_2\), while the marginal risks remain fixed throughout calibration. 

We generate a Monte Carlo reference sample of size \(N=200{,}000\) from the specified covariate distribution and apply Algorithm~\ref{alg:structure-preserving-calibration}. 
The calibration path is evaluated on \(K=100\) equally spaced control-arm relative marginal prognostic strength values spanning \(95\%\) of its attainable range. 
At each point, we evaluate the treatment-arm relative marginal prognostic strength and, under equal randomization, the relative precision-gain potential for the marginal risk difference, risk ratio, and odds ratio. 

\begin{figure}[ht] 
\centering 
\includegraphics[width=1.0\linewidth]{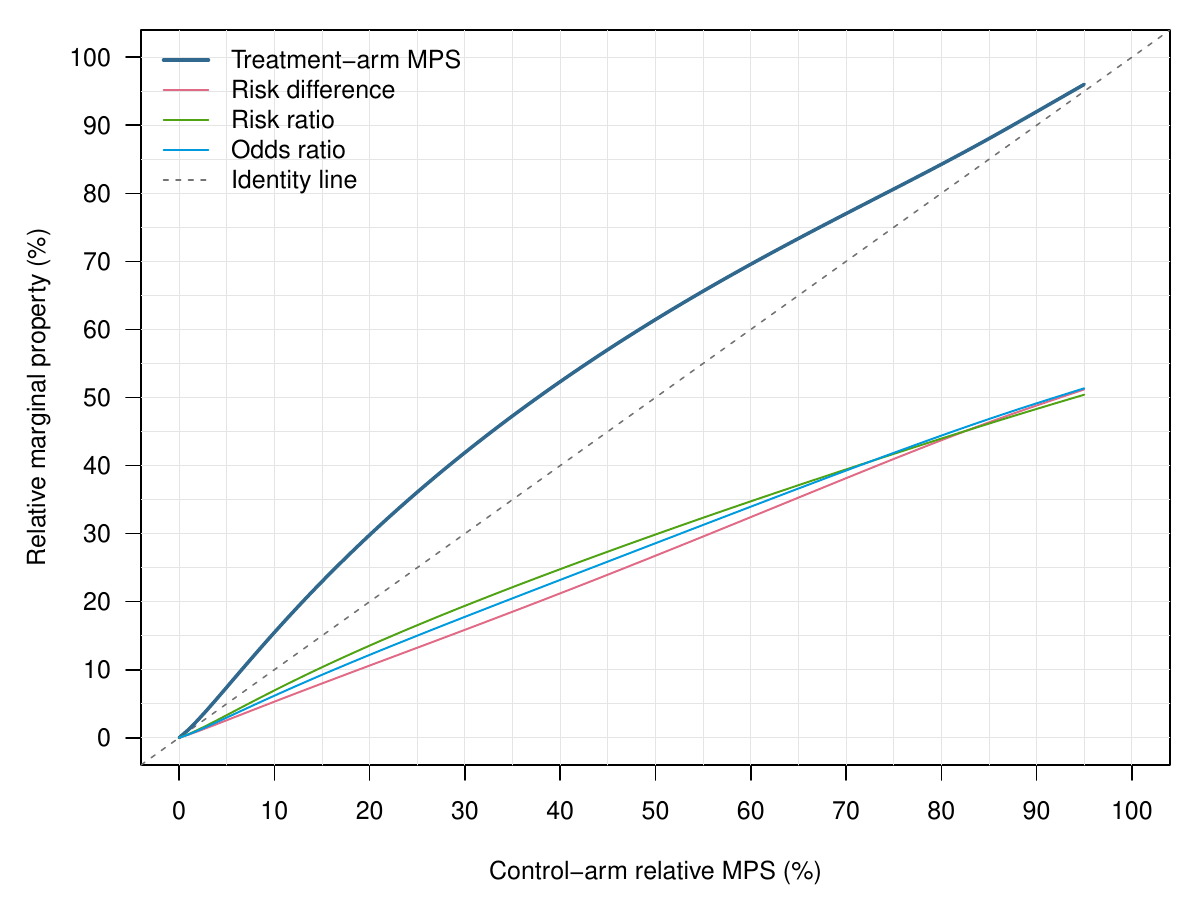} \caption{ Structure-preserving calibration path under equal randomization. Treatment-arm relative marginal prognostic strength and relative precision-gain potential for the marginal risk difference, risk ratio, and odds ratio are shown as functions of control-arm relative marginal prognostic strength. } \label{fig:calibration-path} 
\end{figure} 

Figure~\ref{fig:calibration-path} displays the resulting calibration path. 
Each value of the control-arm marginal prognostic strength identifies a unique structure-preserving data-generating mechanism and is associated with a corresponding treatment-arm marginal prognostic strength and estimand-specific precision-gain potential. 
The separation of the precision-gain curves illustrates that data-generating mechanisms with the same qualitative coefficient structure, marginal risks, and arm-specific prognostic strengths can imply different attainable precision gains for different marginal estimands. 

To obtain a specific simulation scenario, we select a control-arm relative marginal prognostic strength of $20\%$. 
Rather than selecting the closest point on the display grid, the algorithm recalibrates the data-generating mechanism directly to this target. 
The resulting common multiplier is $\lambda^\star=2.0932$, with calibrated intercepts 
\[ \alpha_0(\lambda^\star)=-0.2378, \qquad \alpha_1(\lambda^\star)=-2.9674. \] 
The calibrated outcome model is therefore 
\[ \begin{aligned} 
\logit\left\{\E(Y\mid A,X)\right\} ={}& -0.2378 + (-2.9674+0.2378)A \\ 
& + 2.0932 \left( 0.60X_1 + 0.40X_2 + 0.25X_1X_2^2 - 1.10AX_2 \right). \end{aligned} \] 
The resulting relative marginal prognostic strengths are 
\[\mathcal S_{0,\lambda^\star}^{\mathrm{rel}}(X) =20.0\%, \qquad \mathcal S_{1,\lambda^\star}^{\mathrm{rel}}(X)=29.8\%. \] 
Under equal randomization, the corresponding relative precision-gain potentials are \(10.6\%\), \(13.5\%\), and \(12.2\%\) for the marginal risk difference, risk ratio, and odds ratio, respectively. 
The application thus illustrates how the proposed algorithm constructs simulation scenarios with controlled population-level prognostic information and explicit estimand-specific precision-gain potential while preserving the prespecified coefficient structure and marginal risks.

\section{Discussion}
\label{sec:discussion}

Covariate adjustment is commonly motivated by the intuition that accounting for prognostic baseline information should improve the precision of treatment-effect estimation. 
For marginal treatment effects with binary endpoints, our results show that this intuition requires qualification. 
The attainable precision gain is not determined by the apparent prognostic strength of the covariates alone, but depends on how the arm-specific risk functions combine for the target estimand and treatment allocation. 
Consequently, covariates with strong conditional associations with the outcome need not provide substantial precision gain, and even substantial population-level prognostic information within both treatment arms need not translate into greater precision for estimation of a marginal treatment effect.

Our first main contribution is a general identity that quantifies the maximal first-order variance reduction attainable through covariate adjustment for a wide range of marginal estimands. 
The identity makes explicit how precision-gain potential depends jointly on the arm-specific risk functions, the target estimand, and treatment allocation. 
It generalizes an earlier result of \citet{MooreVanderLaan2009} for the marginal risk difference under equal randomization to arbitrary differentiable marginal estimands and treatment allocation probabilities. 
Building on this identity, we introduced marginal prognostic strength as an arm-specific, population-level measure of the prognostic information carried by the adjustment covariates. 
This formalization shows why the magnitude of conditional outcome-model coefficients does not by itself characterize prognostic strength at the population level. 
It also clarifies why even marginal prognostic strength does not generally determine precision-gain potential.
When the covariates are non-predictive on the risk-difference or risk-ratio scale, the usual intuition that greater prognostic strength leads to greater precision holds.
Outside these settings, however, the contributions from the two treatment arms may offset one another, so that substantial prognostic strength need not translate into substantial precision gain for the treatment effect.

Our second main contribution is a structure-preserving calibration algorithm for methodological simulation studies.
Rather than relying on coefficient magnitude as a proxy for how informative a simulation scenario is for studying covariate adjustment, the procedure preserves a user-specified qualitative conditional outcome structure and marginal risks while calibrating the data-generating mechanism on an interpretable scale of marginal prognostic strength. 
Precision-gain potential can then be evaluated explicitly for the estimand and treatment allocation of interest. 
The proposed calibration framework therefore provides a principled way to construct simulation scenarios with controlled marginal properties and known potential for efficiency gains.

Some limitations of our work deserve consideration, along with potential avenues for future work. 
First, precision-gain potential is a first-order asymptotic quantity and therefore does not provide a finite-sample guarantee for the precision gain achieved by a particular adjusted estimator. 
Finite-sample performance additionally depends on the estimator, estimation of nuisance functions, regularization, and other implementation choices.
Nevertheless, the framework remains informative for both methodological understanding and simulation design.
The precision-gain identity makes explicit the factors that determine the attainable efficiency gain and how they interact, while the calibration procedure allows simulation scenarios to be constructed with a prespecified asymptotic potential for precision gain.
It therefore provides a principled benchmark against which the finite-sample behavior of adjustment methods can be studied.

Second, the structure-preserving calibration algorithm developed here is based on logistic outcome models and a common multiplier applied to the conditional log-odds coefficients. 
This restriction concerns the proposed calibration procedure rather than the underlying precision-gain framework. 
Alternative outcome models or alternative paths through the space of data-generating mechanisms may be preferable when different aspects of the conditional outcome structure are to be preserved. 
Developing and formally characterizing such extensions represents an important direction for future methodological work.

Finally, although the precision-gain identity itself may extend beyond binary endpoints, the present development of marginal prognostic strength and, in particular, the structure-preserving calibration algorithm is tailored to binary outcomes.
Similar issues are likely to arise for other endpoint types, such as time-to-event endpoints, whenever covariate adjustment relies on nonlinear models and marginal treatment effects are obtained from conditional outcome models. 
In addition, the framework presented here has been developed for settings with complete data.
Extending the framework and corresponding calibration methods to other endpoint types and settings with missing data is therefore an important direction for future research.

Overall, our results shift the question from whether a baseline covariate is ``prognostic'' to how the prognostic information carried by the adjustment covariates interacts with the target estimand and trial design. 
The precision-gain identity provides a formal characterization of this relationship, while the calibration algorithm makes it possible to incorporate that characterization directly into the design of methodological simulation studies. 
Together, these results provide a framework for understanding, quantifying, and deliberately controlling the potential precision gains from covariate adjustment for binary endpoints.

\vspace*{1cm}
\bibliographystyle{plainnat} 
\bibliography{references}

\newpage
\clearpage
\appendix
\section*{Appendix}
\addcontentsline{toc}{section}{Appendix}
\setcounter{coro}{0}
\renewcommand{\thecoro}{A\arabic{coro}}
\setcounter{figure}{0}
\renewcommand{\thefigure}{A\arabic{figure}}
\section{Precision-Gain Potential}
\subsection{Proof of Proposition 1 (Precision-Gain Identity for Differentiable Marginal Estimands)}

\begin{proof}
The proof combines two standard ingredients. First, we use the influence functions for treatment-specific marginal risks under unadjusted and under efficient covariate-adjusted estimation in the nonparametric model under consideration. 
Second, we apply the multivariate delta method to the differentiable map \(f(\mu_1,\mu_0)\).

The unadjusted estimators of \(\mu_1\) and \(\mu_0\) are the arm-specific sample means with influence functions
\[
\phi_{1,\mathrm{unadj}}
=
\frac{A}{\pi}(Y-\mu_1),
\qquad
\phi_{0,\mathrm{unadj}}
=
\frac{1-A}{1-\pi}(Y-\mu_0).
\]
For semiparametrically efficient covariate-adjusted estimators, the influence functions are 
\[
\phi_{1,\mathrm{eff}}
=
p_1(X)-\mu_1+\frac{A}{\pi}(Y-p_1(X)),
\qquad
\phi_{0,\mathrm{eff}}
=
p_0(X)-\mu_0+\frac{1-A}{1-\pi}(Y-p_0(X)).
\]

Let \(s\in\{\mathrm{unadj},\mathrm{eff}\}\) denote either analysis strategy, and let $\hat \mu_{a,s}$ denote the corresponding estimator of $\mu_a$ with influence function $\phi_{a,s}$. By the multivariate delta method,
\[
\sqrt n\bigl(f(\hat\mu_{1,s},\hat\mu_{0,s})-f(\mu_1,\mu_0)\bigr)
\overset{d}{\longrightarrow}
\mathcal N\bigl(0,\nu_s(f)\bigr),
\]
where the first-order variance constant is
\[
\nu_s(f)
=
f_1^2\Var(\phi_{1,s})
+
2f_1f_0\Cov(\phi_{1,s},\phi_{0,s})
+
f_0^2\Var(\phi_{0,s}).
\]

For the unadjusted estimator,
\[
\nu_{\mathrm{unadj}}(f)
=
f_1^2\frac{\Var(Y^1)}{\pi}
+
f_0^2\frac{\Var(Y^0)}{1-\pi}, 
\]
because
\begin{align*}
    \Var(\phi_{1,\mathrm{unadj}})
    &=\E[\phi_{1,\mathrm{unadj}}^2]-\E[\phi_{1,\mathrm{unadj}}]^2 && \\
    &= \pi^{-2}\E[\{A(Y-\mu_1)\}^2] && \weil{since $\E[\phi_{1,\mathrm{unadj}}]=0$}\\
    &= \pi^{-2}\E\{A(Y^1-\mu_1)^2\} && \weil{by consistency and $A^2=A$}\\
    &= \pi^{-1}\E\{(Y^1-\mu_1)^2\} && \weil{since $A\indep Y^1$}\\
    &= \pi^{-1}\Var(Y^1)
\end{align*}
and analogously
\begin{align*}
    \Var(\phi_{0,\mathrm{unadj}})=(1-\pi)^{-1}\Var(Y^0)
\end{align*}
and finally 
\begin{align*}
    \Cov(\phi_{1,\mathrm{unadj}},\phi_{0,\mathrm{unadj}})
    &= \E[\phi_{1,\mathrm{unadj}}\cdot \phi_{0,\mathrm{unadj}}] && \weil{since $\E[\phi_{1,\mathrm{unadj}}]=\E[\phi_{0,\mathrm{unadj}}]=0$}\\
    &= \E\left\{\frac{A}{\pi}(Y-\mu_1)\cdot\frac{1-A}{1-\pi}(Y-\mu_0)\right\} && \\
    &=  0 && \weil{since $A(1-A)=0$}
\end{align*}

For the efficient covariate-adjusted estimator,
\[
\nu_{\mathrm{eff}}(f)
=
\Var\{f_1p_1(X)+f_0p_0(X)\}
+
f_1^2\frac{\E\{p_1(X)(1-p_1(X))\}}{\pi}
+
f_0^2\frac{\E\{p_0(X)(1-p_0(X))\}}{1-\pi},
\]
because $\E(\phi_{1,\mathrm{eff}})=\E(\phi_{0,\mathrm{eff}})=0$ and thus
\begin{align*}
    &\Var(\phi_{1,\mathrm{eff}})\\
    &= \E[\phi_{1,\mathrm{eff}}^2]-\E[\phi_{1,\mathrm{eff}}]^2 && \\
    &=\E[\{p_1(X)-\mu_1+\pi^{-1}A(Y-p_1(X))\}^2] && \weil{since $\E[\phi_{1,\mathrm{eff}}]=0$}\\
    &=\begin{aligned}[t]&\E\{(p_1(X)-\mu_1)^2\}\\&+\E\{2(p_1(X)-\mu_1)\pi^{-1}A(Y-p_1(X))\}\\&+\E\{\pi^{-2}A^2(Y-p_1(X))^2\}\end{aligned} && \\
    &=\begin{aligned}[t]&\Var\{p_1(X)\}\\&+2\pi^{-1}\E\{(p_1(X)-\mu_1)A(Y^1-p_1(X))\}\\&+\pi^{-2}\E\{A(Y^1-p_1(X))^2\}\end{aligned} && \weil{by consistency and $A^2=A$}\\
    &=\begin{aligned}[t]&\Var\{p_1(X)\}\\&+2\E\{(p_1(X)-\mu_1)(Y^1-p_1(X))\}\\&+\pi^{-1}\E\{(Y^1-p_1(X))^2\}\end{aligned} && \weil{since $(\mathrm{A\indep (Y^1,X)})$}\\
    &=\begin{aligned}[t]&\Var\{p_1(X)\}\\&+2\E\{\E(p_1(X)Y^1-p_1(X)^2-\mu_1Y^1+\mu_1p_1(X)\mid X)\}\\
    &+\pi^{-1}\E\{\E(Y^1-2Y^1p_1(X)+p_1(X)^2 \mid X)\}\end{aligned} && \weil{by the tower property}\\
    &= \begin{aligned}[t]&\Var\{p_1(X)\}\\&+2\E\{p_1(X)\E(Y^1\mid X)-p_1(X)^2-\mu_1\E(Y^1\mid X)+\mu_1p_1(X)\}\\
    &+\pi^{-1}\E\{\E(Y^1\mid X)-2p_1(X)\E(Y^1\mid X)+p_1(X)^2\}\end{aligned} && \weil{since $p_1(X)$ is $X$-measurable} \\
    &=\Var\{p_1(X)\}+\pi^{-1}\E\{p_1(X)(1-p_1(X))\}
\end{align*}
and analogously 
\begin{align*}
    \Var(\phi_{0,\text{eff}})=\Var\{p_0(X)\}+\{1-\pi\}^{-1}\E\{p_0(X)(1-p_0(X))\}
\end{align*}
and finally
\begin{align*}
&\Cov(\phi_{1,\mathrm{eff}},\phi_{0,\mathrm{eff}})\\
&=\E\left[\left\{p_1(X)-\mu_1+\frac{A}{\pi}(Y-p_1(X))\right\}\left\{p_0(X)-\mu_0+\frac{1-A}{1-\pi}(Y-p_0(X))\right\}\right] && \\
 &=\E\left[\left\{p_1(X)-\mu_1+\frac{A}{\pi}(Y^1-p_1(X))\right\}\left\{p_0(X)-\mu_0+\frac{1-A}{1-\pi}(Y^0-p_0(X))\right\}\right] && \weil{by consistency}\\
 &=\begin{aligned}[t]&\E\{(p_1(X)-\mu_1)(p_0(X)-\mu_0)\}\\&+\E\left\{(p_1(X)-\mu_1)\frac{1-A}{1-\pi}(Y^0-p_0(X))\right\}\\
 &+\E\left\{(p_0(X)-\mu_0)\frac{A}{\pi}(Y^1-p_1(X))\right\}\\
 &+\E\left\{\frac{A}{\pi}(Y^1-p_1(X))\frac{1-A}{1-\pi}(Y^0-p_0(X))\right\}\end{aligned}\\
 &= \begin{aligned}[t]&\Cov\{p_1(X),p_0(X)\}\\&+\E\left\{(p_1(X)-\mu_1)\frac{1-A}{1-\pi}(Y^0-p_0(X))\right\}\\
 &+\E\left\{(p_0(X)-\mu_0)\frac{A}{\pi}(Y^1-p_1(X))\right\}\end{aligned} && \weil{since $A(1-A)=0$}\\
 &=\begin{aligned}[t]& \Cov\{p_1(X),p_0(X)\}\\&+\E\{(p_1(X)-\mu_1)(Y^0-p_0(X))\}\\ &+\E\{(p_0(X)-\mu_0)(Y^1-p_1(X))\}\end{aligned} && \weil{$A\indep (X,Y^0,Y^1)$}\\
  &=\begin{aligned}[t]&\Cov[p_1(X),p_0(X)]\\&+\E[\E\{(p_1(X)-\mu_1)(Y^0-p_0(X))\mid X\}]\\
  &+\E[\E\{(p_0(X)-\mu_0)(Y^1-p_1(X))\mid  X\}]\end{aligned} && \weil{by the tower property}\\
    &=\begin{aligned}[t]&\Cov[p_1(X),p_0(X)]\\&+\E[\{p_1(X)-\mu_1\}\{\E(Y^0\mid X)-p_0(X)\}]\\
  &+\E[\{p_0(X)-\mu_0\}\{\E(Y^1\mid X)-p_1(X)\}]\end{aligned} && \weeil{$p_{0}(X),p_1(X)$ are}{$X$-measurable}\\
  &=\Cov(p_1(X),p_0(X))
\end{align*}
    
Using the law of total variance,
\[
\Var(Y^a)
=
\Var\{p_a(X)\}
+
\E\{p_a(X)(1-p_a(X))\},
\qquad a\in\{0,1\},
\]
we obtain
\[
\begin{aligned}
\mathcal{G}^\mathrm{abs}_{\pi,f}(X)
&=
\nu_{\mathrm{unadj}}(f)-\nu_{\mathrm{eff}}(f)\\
&=
f_1^2\frac{1-\pi}{\pi}\Var(p_1(X))
+
f_0^2\frac{\pi}{1-\pi}\Var(p_0(X))
-
2f_1f_0\Cov(p_1(X),p_0(X))\\
&=
\Var\left(
\sqrt{\frac{1-\pi}{\pi}}\,f_1p_1(X)
-
\sqrt{\frac{\pi}{1-\pi}}\,f_0p_0(X)
\right)
\end{aligned}
\]
and \[
\begin{aligned}
\mathcal{G}^\mathrm{rel}_{\pi,f}(X)
&=\frac{\Var\left(
\sqrt{\frac{1-\pi}{\pi}}\,f_1p_1(X)
-
\sqrt{\frac{\pi}{1-\pi}}\,f_0p_0(X)
\right)}{f_1^2\frac{\Var(Y^1)}{\pi}
+
f_0^2\frac{\Var(Y^0)}{1-\pi}}
\end{aligned}
\]
which proves Proposition 1.
\end{proof}

\newpage
\subsection{Corollary \ref{coro:reparameterization} (Invariance Under Reparameterization)}
\label{app:proof-cor2}

A further insight from Proposition 1 is that relative precision-gain potential is invariant to differentiable reparameterizations.
\begin{coro}[Invariance Under Reparameterization] 
\label{coro:reparameterization} 
Let $\psi=f(\mu_1,\mu_0)$ satisfy the assumptions of Proposition~1, and let $\widetilde\psi=g(\psi)$, where \(g\) is differentiable at \(\psi\) with \(g'(\psi)\neq0\). 
Then, 
\[
\mathcal G^{\mathrm{abs}}_{\pi,\widetilde\psi}(X) = g'(\psi)^2 \mathcal G^{\mathrm{abs}}_{\pi,\psi}(X) \qquad \text{ and }\qquad  \mathcal G^{\mathrm{rel}}_{\pi,\widetilde\psi}(X) = \mathcal G^{\mathrm{rel}}_{\pi,\psi}(X).
\] 
\end{coro}
\begin{proof}
Applying Corollary 1 to the estimand given by \[
(\mu_1,\mu_0)\mapsto g(f(\mu_1,\mu_0))
\]
yields
\[
\begin{aligned}
  \mathcal{G}^\mathrm{abs}_{\pi, \widetilde{\psi}}(X)&=\Var\left(\sqrt{\frac{1-\pi}{\pi}}\frac{\partial}{\partial \mu_1}g(f(\mu_1,\mu_0))p_1^S(X)-\sqrt{\frac{\pi}{1-\pi}}\frac{\partial }{\partial \mu_0}g(f(\mu_1,\mu_0))p_0^S(X)\right)\\
  &=\Var\left(\sqrt{\frac{1-\pi}{\pi}}g'(f(\mu_1,\mu_0))f_1p_1^S(X)-\sqrt{\frac{\pi}{1-\pi}}g'(f(\mu_1,\mu_0))f_0p_0^S(X)\right)\\
  &=g'(\psi)^2\mathcal{G}^\mathrm{abs}_{\pi,\psi}(X)
\end{aligned}
\]
Applying the unadjusted variance formula derived in the proof of Proposition 1 to the map \(g\circ f\), we obtain \[
\begin{aligned}
\nu_\mathrm{unadj}(g\circ f)&=\left(\frac{\partial }{\partial \mu_1}g(f(\mu_1,\mu_0))\right)^2 \frac{\Var(Y^1)}{\pi}+\left(\frac{\partial}{\partial \mu_0}g(f(\mu_1,\mu_0))\right)^2 \frac{\Var(Y^0)}{1-\pi}\\
&=g'(\psi)^2\left(f_1^2\frac{\Var(Y^1)}{\pi}+f_0^2\frac{\Var(Y^0)}{1-\pi}\right)\\
&=g'(\psi)^2 \nu_\mathrm{unadj}(f).
\end{aligned}
\]
Consequently,
\[
\mathcal{G}^\mathrm{rel}_{\pi,\widetilde{\psi}}(X)=\frac{g'(\psi)^2\mathcal{G}^\mathrm{abs}_{\pi,\psi}(X)}{g'(\psi)^2 \nu_\mathrm{unadj}(f)}=\mathcal{G}^\mathrm{rel}_{\pi,\psi}(X),
\]
which proves the corollary.
\end{proof}
Thus, for example, the risk ratio and log risk ratio have the same relative precision-gain potential, as do the odds ratio and log odds ratio.

\newpage
\section{From Prognostic Strength to Precision Gain}
\subsection{Proof of Corollary 1 (Precision-Gain Under Non-Predictiveness)}
\label{app:proof-cor-nonpredictive}
\begin{proof}
Under non-predictiveness on the target effect scale, there exists a deterministic relationship between the arm-specific risk functions.

For the risk difference,
\[
p_1^S(X)
=
p_0^S(X)+\Delta ,
\]
where $\Delta$ is the constant conditional risk difference.
For the risk ratio,
\[
p_1^S(X)
=
\rho\,p_0^S(X),
\]
where $\rho=\mu_1/\mu_0$ is the constant conditional risk ratio.

\paragraph{Risk difference}
For
$f(\mu_1,\mu_0)=\mu_1-\mu_0$,
the gradient is
\[
(f_1,f_0)=(1,-1).
\]
Substituting
$p_1^S(X)=p_0^S(X)+\Delta$
into Corollary 1 gives
\[
\begin{aligned}
\mathcal{G}^\mathrm{abs}_{\pi,\mathrm{RD}}(X)
&=
\Var\!\left\{
\sqrt{\frac{1-\pi}{\pi}}
(p_0^S(X)+\Delta)
+
\sqrt{\frac{\pi}{1-\pi}}
p_0^S(X)
\right\}
\\
&=
\left(
\sqrt{\frac{1-\pi}{\pi}}
+
\sqrt{\frac{\pi}{1-\pi}}
\right)^2
\Var\!\left(
p_0^S(X)
\right)\\
&=\frac{1}{\pi(1-\pi)}
\Var\!\left(
p_0^S(X)
\right).
\end{aligned}
\]

Because adding a constant does not change variance,
\[
\Var\!\left(
p_1^S(X)
\right)
=
\Var\!\left(
p_0^S(X)
\right),
\]
and therefore
\[
\mathcal{G}^\mathrm{abs}_{\pi,\mathrm{RD}}(X)
=
\frac{1}{\pi(1-\pi)}
\mathcal{S}^\mathrm{abs}_{0}(X)
=
\frac{1}{\pi(1-\pi)}
\mathcal{S}^\mathrm{abs}_{1}(X).
\]
This proves Assertion~(i).

\paragraph{Risk ratio}

For
$f(\mu_1,\mu_0)=\mu_1/\mu_0$,
\[
(f_1,f_0)
=
\left(
\frac{1}{\mu_0},
-\frac{\rho}{\mu_0}
\right),
\]
and
$p_1^S(X)=\rho\,p_0^S(X)$.

Substituting into Corollary 1 yields
\[
\begin{aligned}
\mathcal{G}^\mathrm{abs}_{\pi,\mathrm{RR}}(X)
&=
\Var\!\left\{
\frac{\rho}{\mu_0}
\left(
\sqrt{\frac{1-\pi}{\pi}}
+
\sqrt{\frac{\pi}{1-\pi}}
\right)
p_0^S(X)
\right\}
\\
&=
\frac{\rho^2}
{\mu_0^2\,\pi(1-\pi)}
\Var\!\left(
p_0^S(X)
\right).
\end{aligned}
\]

Since
\[
\Var\!\left(
p_1^S(X)
\right)
=
\rho^2
\Var\!\left(
p_0^S(X)
\right),
\]
this implies
\[
\mathcal{G}^\mathrm{abs}_{\pi,\mathrm{RR}}(X)
=
\frac{\rho^2}
{\mu_0^2\,\pi(1-\pi)}
\mathcal{S}^\mathrm{abs}_{0}(X)
=
\frac{1}
{\mu_0^2\,\pi(1-\pi)}
\mathcal{S}^\mathrm{abs}_{1}(X).
\]
This proves Assertion~(ii).

\paragraph{Relative precision gain}

Divide the equations from Assertions~(i) and~(ii) by the corresponding unadjusted variance
\[
\nu_{\mathrm{unadj}}(f)
=
\frac{f_1^2\Var(Y^1)}{\pi}
+
\frac{f_0^2\Var(Y^0)}{1-\pi}, \qquad f\in\{\RD, \RR\}.
\]
Under non-predictiveness on the risk-difference scale, $\mathcal S^\mathrm{abs}_{0}(X) = \mathcal S^\mathrm{abs}_{1}(X)$, whereas under non-predictiveness on the risk-ratio scale, $\rho^2\mathcal S^\mathrm{abs}_{0}(X) = \mathcal S^\mathrm{abs}_{1}(X)$. Using these relationships together with the definition $\mathcal S^\mathrm{abs}_{a}(X) = \mathcal S^\mathrm{rel}_{a}(X)\Var(Y^a)$ for $a\in\{0,1\}$ yields, in both cases,
\[
\mathcal{G}^\mathrm{rel}_{\pi,f}(X)
=
\left(
\frac{1-\pi}
     {\mathcal{S}^{\mathrm{rel}}_{1}(X)}
+
\frac{\pi}
     {\mathcal{S}^\mathrm{rel}_{0}(X)}
\right)^{-1},
\qquad
f\in\{\mathrm{RD},\mathrm{RR}\}.
\]
This proves Assertion~(iii).
\end{proof}
\subsection{Counterexamples for the Odds Ratio}
\label{app:mlor-counterexamples}
In this section, we provide examples illustrating that, for the marginal odds ratio, non-predictiveness of the covariates induces a nonlinear relationship between the arm-specific risk functions, such that marginal prognostic strength does not generally determine precision-gain potential. 
For computational convenience, we formulate the examples on the log odds-ratio (LOR) scale. 
By Corollary~\ref{coro:reparameterization}, relative precision-gain potential is invariant under this reparameterization, so the conclusions apply equally to the odds ratio.

\paragraph{Absolute quantities}
This example shows that (absolute) marginal prognostic strength does not determine (absolute) precision-gain potential. 
We compare the two data-generating mechanisms described by the covariate $X\sim\bernoulli(1/2)$, the randomization probability $\pi=1/2$ and the conditional risks given in the first two rows of Table~\ref{tab:mlor-counterexamples}.
\begin{table}[ht]
\centering
\begin{threeparttable}
\captionsetup{justification=centering}
\caption{Data-generating mechanisms for the LOR counterexamples}
\label{tab:mlor-counterexamples}
\begin{tabular}{l|cccc}
\toprule
DGM &$p_0(0)$& $p_0(1)$ & $p_1(0)$ & $p_1(1)$ \\
\midrule
I& 7/52 & 3/8 & 1/10 & 3/10 \\

II & 5/13 & 5/8 & 1/5 & 2/5 \\

III& 1/2 & 4/5 & 1/5 & 1/2 \\

IV & 1/2 & 4/5 & 2/5 & 8/11 \\
\bottomrule
\end{tabular}
\end{threeparttable}
\end{table}
It holds that \[
\log\left(\frac{p_1(X)/(1-p_1(X))}{p_0(X)/(1-p_0(X))}\right)=\begin{cases}
    \log(5/7)\qquad \text{for both levels of $X$ in DGM I}\\
    \log(2/5)\qquad \text{for both levels of $X$ in DGM II.}
\end{cases}
\] Thus, $X$ is non-predictive on the LOR scale in both data-generating mechanisms under consideration.
Using \[
\Var(p_a(X))=\Var(X)\cdot (p_a(1)-p_a(0))^2,
\]
we obtain 
\[
\mathcal{S}^\mathrm{abs}_{a}(X)=\begin{cases}
   1/100\qquad \qquad \text{for $a=1$ in DGM I and II}\\
    625/43264\qquad \text{for $a=0$ in DGM I and II.}
\end{cases}
\]
Hence, the arm-specific marginal prognostic strength of $X$ is the same in both data-generating mechanisms. 
The precision-gain identity for the LOR and $\pi=1/2$ reduces to 
\[
\mathcal{G}^\mathrm{abs}_{1/2,\mathrm{LOR}}(X)=\Var\left\{\frac{p_1(X)}{\mu_1(1-\mu_1)}+\frac{p_0(X)}{\mu_0(1-\mu_0)}\right\},
\]
yielding 
\[
\mathcal{G}^\mathrm{abs}_{1/2,\mathrm{LOR}}(X)=\begin{cases}
    \frac{273406225}{172764736}\approx 1.583\qquad \text{for DGM I}\\
    \frac{476100}{519841}\approx 0.916\qquad\quad \text{for DGM II.}
\end{cases}
\] Consequently, the same arm-specific marginal prognostic strength can result in different amounts of precision-gain potential for the LOR, i.e. the former quantity does not determine the latter.
\paragraph{Relative quantities}
We now show that relative marginal prognostic strength does not determine relative precision-gain potential. Let $X\sim \bernoulli(1/3)$, $\pi=1/2$ and consider the last two data-generating mechanisms described by the conditional risks displayed in Table~\ref{tab:mlor-counterexamples}.

We have \[
\log\left(\frac{p_1(X)/(1-p_1(X))}{p_0(X)/(1-p_0(X))}\right)=\begin{cases}
    \log(1/4)\qquad \text{for both levels of $X$ in DGM III}\\
    \log(2/3)\qquad \text{for both levels of $X$ in DGM IV,}
\end{cases}
\]
that is, $X$ is non-predictive on the LOR scale in both data-generating mechanisms. 
The arm-specific relative marginal prognostic strengths are given by 
\[
\mathcal{S}^\mathrm{rel}_{a}(X)=\begin{cases}
    2/21\qquad \text{for $a=1$ in DGM III and IV}\\
    1/12\qquad \text{for $a=0$ in DGM III and IV.}
\end{cases}
\]
Thus, the arm-specific relative marginal prognostic strength of $X$ is the same in both data-generating mechanisms. 
However,
\[
\mathcal{G}^\mathrm{rel}_{1/2,\mathrm{LOR}}(X)=\begin{cases}
    \frac{5}{56}\approx 0.08929\qquad \quad\text{for DGM III}\\
    \frac{1849}{20748}\approx 0.08912\qquad\text{for DGM IV,} 
\end{cases}
\]
implying that relative marginal prognostic strength does not determine relative precision-gain potential for the LOR.

\newpage
\section{Structure-Preserving Calibration of Data-Generating Mechanisms}
\subsection{Proof of Proposition 2 (Monotonicity of Marginal Prognostic Strength)}
\begin{proof}
\label{app:proof-monotonicity}
We show that 
\[
\frac{d}{d\lambda}\Var(p_{a,\lambda}(X))>0 \text{ on } (0,\infty).
\]
We start by proving that $\expit^2(\alpha_a(\lambda)+\lambda\eta_a(X))$ is almost surely continuously differentiable with respect to $\lambda$. Define 
\[
F:\R^2 \to \R, \quad (\tilde{\lambda},\alpha)\mapsto \E(\expit(\alpha+\tilde{\lambda} \eta_a(X)))-\mu_a.
\]
Note that, as argued before, for every $\tilde{\lambda}\in\R$, $F(\tilde{\lambda},\alpha)=0$ has a unique solution given by $\alpha_a(\tilde{\lambda})$. 
The function $F$ is therefore an implicit representation of the function $\alpha_a(\tilde{\lambda})$.
To apply the implicit function theorem, we have to show that $F\in C^1(\R^2,\R)$. 
The $\expit$-function is in $C^1$, its derivative takes values in $(0,1/4]$ and $\eta_a(X)$ is integrable.
Thus, dominated convergence yields
\begin{align*}
    \frac{\partial}{\partial \tilde{\lambda}}F(\tilde{\lambda},\alpha)
    &=\lim_{\delta\to 0}\frac{F(\tilde\lambda+\delta,\alpha)-F(\tilde{\lambda},\alpha)}{\delta}\\
    &=\lim_{\delta \to 0}\E\left\{\frac{\expit(\alpha+(\tilde{\lambda}+\delta)\eta_a(X))-\expit(\alpha+\tilde{\lambda}\eta_a(X))}{\delta}\right\}\\
     &=\E\left\{\lim_{\delta \to 0}\frac{\expit(\alpha+(\tilde{\lambda}+\delta)\eta_a(X))-\expit(\alpha+\tilde{\lambda}\eta_a(X))}{\delta}\right\}\\
    &=\E(\eta_a(X)\expit'(\alpha+\tilde{\lambda}\eta_a(X)))
\end{align*}
and
\begin{align*}
    \frac{\partial}{\partial\alpha}F(\tilde{\lambda},\alpha)
    &=\lim_{\delta\to 0}\frac{F(\tilde{\lambda},\alpha+\delta)-F(\tilde{\lambda},\alpha)}{\delta}\\
    &=\lim_{\delta\to 0}\E\left(\frac{\expit(\alpha+\delta+\tilde{\lambda}\eta_a(X))-\expit(\alpha+\tilde{\lambda}\eta_a(X))}{\delta}\right)\\
     &=\E\left(\lim_{\delta\to 0}\frac{\expit(\alpha+\delta+\tilde{\lambda}\eta_a(X))-\expit(\alpha+\tilde{\lambda}\eta_a(X))}{\delta}\right)\\
    &=\E(\expit'(\alpha+\tilde{\lambda}\eta_a(X)))
\end{align*}
where, by Lagrange's mean value theorem, $1+|\eta_a(X)|$ can be used as the majorant in both cases.
Because $\expit'$ is continuous, the same strategy yields that both partial derivatives are continuous. 
Thus, $F\in C^1$.
Furthermore, 
\[
\frac{\partial}{\partial \alpha}F(\tilde{\lambda},\alpha)>0\text{ on }\R^2.
\]
Hence, for every fixed $\lambda\in\R$, the implicit function theorem applied at $(\lambda,\alpha_a(\lambda))$ yields a locally defined continuously differentiable function $\tilde{\lambda}\mapsto \tilde{\alpha}(\tilde{\lambda})$ satisfying $F(\tilde{\lambda},\tilde{\alpha}(\tilde{\lambda}))=0$. By uniqueness of the solution to $F(\tilde{\lambda},\alpha)=0$ in $\alpha$ for each fixed $\tilde{\lambda}$, this local function coincides with the function $\tilde{\lambda}\mapsto \alpha_a(\tilde{\lambda})$. Since this reasoning applies for every $\lambda\in\R$, it follows that $\alpha_a$ is continuously differentiable everywhere, and that 
\[
\alpha_a'(\tilde{\lambda})=-\frac{\E(\eta_a(X)\expit'(\alpha_a(\tilde{\lambda})+\tilde{\lambda}\eta_a(X)))}{\E(\expit'(\alpha_a(\tilde{\lambda})+\tilde{\lambda}\eta_a(X)))}.
\] 
Integrability of $\eta_a(X)$ implies $|\eta_a(X)|<\infty$ a.s., and thus also the composition $\lambda\mapsto \expit^2(\alpha_a(\lambda)+\lambda\eta_a(X))$ is a.s. continuously differentiable. 

Next, we want to interchange differentiation and integration to conclude
\[
\frac{d}{d\lambda}\E(p_{a,\lambda}^2(X))=\E\left(\frac{\partial}{\partial \lambda}p_{a,\lambda}^2(X)\right).
\]
For this, let $\lambda\in\R$ be fixed and consider the restriction of the random function $\tilde{\lambda} \mapsto \expit^2(\alpha_a(\tilde{\lambda})+\tilde{\lambda}\eta_a(X))$ to $(\lambda-1,\lambda+1)$. It holds that 
\[
\E\left(\sup_{\tilde{\lambda}\in (\lambda-1,\lambda+1)}\expit^2(\alpha_a(\tilde{\lambda})+\tilde{\lambda}\eta_a(X))\right)\leq 1<\infty,
\] 
the restricted function is almost surely continuously differentiable, and 
\begin{align*}
&\E\left(\sup_{\tilde{\lambda}\in (\lambda-1,\lambda+1)} \left|\frac{\partial}{\partial \tilde{\lambda}}\expit^2(\alpha_a(\tilde{\lambda})+\tilde{\lambda}\eta_a(X))\right|\right)\\
&=\E\left\{ \sup_{\tilde{\lambda}\in{(\lambda-1,\lambda+1)}}|2\expit(\alpha_a(\tilde{\lambda})+\tilde{\lambda}\eta_a(X))\expit'(\alpha_a(\tilde{\lambda})+\tilde{\lambda}\eta_a(X))(\alpha_a'(\tilde{\lambda})+\eta_a(X))|\right\}\\
&\leq \sup_{\tilde{\lambda}\in{(\lambda-1,\lambda+1)}}|\alpha_a'(\tilde{\lambda})|+\E(|\eta_a(X)|)<\infty,
\end{align*}
where the last inequality holds because the continuous function $\alpha_a'$ is bounded on the compact interval $[\lambda-1,\lambda+1]$ and $\eta_a(X)$ is integrable. Thus, by the Leibniz integral rule,
\[
\frac{d}{d\tilde{\lambda}}\E(p_{a,\tilde{\lambda}}^2(X))=\E\left(\frac{\partial}{\partial \tilde{\lambda}}p_{a,\tilde{\lambda}}^2(X)\right)\quad \forall\tilde{\lambda}\in(\lambda-1,\lambda+1)
\]
and since $\lambda$ was chosen arbitrarily, the identity holds for all $\tilde{\lambda}\in\R$. In particular, $\lambda\mapsto \Var(p_{a,\lambda}(X))$ is differentiable on $\R$.

From the above equation we obtain 
\begin{align*}
   & \frac{d}{d\lambda}\Var(p_{a,\lambda}(X))\\
    &=\E\left(\frac{\partial}{\partial \lambda}p_{a,\lambda}^2(X)\right)\\
    &=\E\{2p_{a,\lambda}(X)\expit'(\alpha_a(\lambda)+\lambda\eta_a(X)) (\eta_a(X)+\alpha_a'(\lambda))\}\\
    &=\E\left[2p_{a,\lambda}(X)\expit'(\alpha_a(\lambda)+\lambda\eta_a(X)) \left\{\eta_a(X)-\frac{\E(\eta_a(X)\expit'(\alpha_a(\lambda)+\lambda\eta_a(X)))}{\E(\expit'(\alpha_a(\lambda)+\lambda\eta_a(X)) )}\right\}\right]\\
    &=\frac{\E\{\expit'(\alpha_a(\lambda)+\lambda\eta_a(X)) \expit'(\alpha_a(\lambda)+\lambda\eta_a(\tilde{X})) (p_{a,\lambda}(X)-p_{a,\lambda}(\tilde X))(\eta_a(X)-\eta_a(\tilde X))\}}{\E\{\expit'(\alpha_a(\lambda)+\lambda\eta_a(X))\}}
\end{align*}
where $\tilde X$ is an independent copy of $X$. 
This expression is strictly positive for $\lambda>0$, because firstly, the function $\eta\mapsto \expit(\alpha_a(\lambda)+\lambda\eta)$ is strictly increasing for $\lambda>0$, secondly $\P(\eta_a(X)\neq \eta_a(\tilde X))>0$ as long as $\eta_a(X)$ is not almost surely constant, and thirdly $\expit'>0$. 
Since $\lambda\mapsto \Var(p_{a,\lambda}(X))$ is differentiable on $\R$ and thus continuous on $\R$, this yields that $\lambda\mapsto \Var(p_{a,\lambda}(X))$ is strictly increasing on $[0,\infty)$. 
Dividing by the constant $\mu_a(1-\mu_a)$ completes the proof.
\end{proof}

\newpage
\subsection{Proof of Proposition 3 (Limiting Marginal Prognostic Strength)}
\label{app:proof-mps-limit}
\begin{proof}
Define 
\[
\mathcal C=\{c\in \R: \P(\eta_a(X)>c)\leq \mu_a\leq \P(\eta_a(X)\geq c)\}.
\]
We first show that $\mathcal C$ is sequentially closed and hence closed. Let $(c_n)_{n\in \mathbb N}$ be a convergent sequence in $\mathcal C$ with limit $c$. We show that $c\in \mathcal C$. Let $\epsilon >0$. By convergence, 
\[
\exists n\in \mathbb N:|c_n-c|<\epsilon.
\]
For such $n$, 
\[
\P(\eta_a(X)>c+\epsilon)\leq \P(\eta_a(X)>c_n)\leq \mu \leq \P(\eta_a(X)\geq c_n)\leq \P(\eta_a(X)>c-\epsilon),
\]
where the second and third inequalities follow because $c_n\in \mathcal C$. 
Now take a decreasing sequence $(\epsilon_m)_{m\in\mathbb N}$ with $\epsilon_m\downarrow 0$. Then sigma continuity of $\P$ implies \[
\P(\eta_a(X)>c)=\P\left(\bigcup_{m\in \mathbb N}\{\eta_a(X)>c+\epsilon_m\}\right)=\lim_{m\to\infty}\P(\eta_a(X)>c+\epsilon_m)\leq \mu.
\]
and 
\[
\mu\leq \lim_{m\to\infty}\P(\eta_a(X)>c-\epsilon_m)=\P\left(\bigcap_{m\in \mathbb N}\{\eta_a(X)>c-\epsilon_m\}\right)=\P(\eta_a(X)\geq c).
\]
Thus, $c\in \mathcal C$, which proves that $\mathcal C$ is closed. We can therefore define $c_\mathrm{min}=\min\mathcal C$ and $c_\mathrm{max}=\max \mathcal C$.

The conditional risk function $p_{a,\lambda_a}$ can be rewritten as 
\[
p_{a,\lambda_a}(X)=\expit(\alpha_a(\lambda_a)+\lambda_a\eta_a(X))=\expit(\lambda_a(\eta_a(X)-c_{\lambda_a})),\qquad c_{\lambda_a}=-\frac{\alpha_a(\lambda_a)}{\lambda_a}.
\]
We now prove that 
\[
\liminf_{\lambda_a\to\infty}c_{\lambda_a}\geq c_\mathrm{min}.
\]
Suppose $\liminf_{\lambda_a\to\infty}c_{\lambda_a}<c_\mathrm{min}$. Then there exists $\delta>0$ and a sequence $(\lambda_n)_{n\in \mathbb N}$ s.t. $\lambda_n\to\infty$ and $\forall n\in\mathbb N:c_{\lambda_n}\leq c_\mathrm{min}-\delta$, where $c_{\lambda_n}=-\alpha_a(\lambda_n)/\lambda_n$. Define $t=c_\mathrm{min}-\frac{\delta}{2}$. From $t<c_\mathrm{min}=\min \mathcal C$, it follows that $\P(\eta_a(X)> t)>\mu_a$ or $\mu_a>\P(\eta_a(X)\geq t)$. But as $t<c_\mathrm{min}$, we have $\P(\eta_a(X)\geq t)\geq \P(\eta_a(X)\geq c_\mathrm{min})\geq \mu_a$ and therefore $\mu_a > \P(\eta_a(X)\geq t)$ cannot hold. Consequently, $\P(\eta_a(X)>t)>\mu_a$ holds. As $\alpha_a(\lambda_a)$ is, by definition, always chosen such that $\mu_a=\E(p_{a,\lambda_a}(X))$, we have 

\begin{align*}
\mu_a
&=\E(\expit(\alpha_a(\lambda_n)+\lambda_n\eta_a(X)))\\
&=\E\{\expit(\lambda_n(\eta_a(X)-c_{\lambda_n}))\}\\
&\geq \E\{\expit(\lambda_n(\eta_a(X)-c_{\lambda_n}))\mathbf 1_{\{\eta_a(X)>t\}}\}\\
&\geq \E\{\expit(\lambda _n(\delta/2))\mathbf 1_{\{\eta_a(X)>t\}}\}\\
&=\expit(\lambda_n\delta/2)\P(\eta_a(X)>t)\to \P(\eta_a(X)>t) \quad (n\to\infty).
\end{align*}
This implies $\mu_a >\mu_a$, which is a contradiction. Thus, 
\[
\liminf_{\lambda_a\to\infty}c_{\lambda_a}\geq c_\mathrm{min}.
\]
Now we prove 
\[
\limsup_{\lambda_a\to\infty}c_{\lambda_a}\leq c_\mathrm{max}.
\]
Suppose $\limsup_{\lambda_a\to\infty}c_{\lambda_a}>c_\mathrm{max}$. Then there exists $\delta >0$ and a sequence $(\lambda_n)_{n\in\mathbb N}$ s.t. $\lambda_n\to\infty$ and $\forall n\in\mathbb N:c_{\lambda_n}\geq c_\mathrm{max}+\delta$. Define $t=c_\mathrm{max}+\frac{\delta}{2}$. From $t>c_\mathrm{max}=\max \mathcal C$, it follows that $\P(\eta_a(X)>t)>\mu_a$ or $\mu_a>\P(\eta_a(X)\geq t)$. But as $t>c_\mathrm{max}$, we have $\P(\eta_a(X)>t)\leq \P(\eta_a(X)>c_\mathrm{max})\leq \mu_a$ and therefore $\P(\eta_a(X)>t)>\mu_a$ cannot hold. Consequently, $\mu_a>\P(\eta_a(X)\geq t)$ holds. Thus,
\begin{align*}
\mu_a
&=\E\{\expit(\lambda_n(\eta_a(X)-c_{\lambda_n}))\}\\
&=\E\{\expit(\lambda_n(\eta_a(X)-c_{\lambda_n}))\mathbf 1_{\{\eta_a(X)\geq t\}}\}+\E\{\expit(\lambda_n(\eta_a(X)-c_{\lambda_n}))\mathbf 1_{\{\eta_a(X)<t\}}\}\\
&\leq \P(\eta_a(X)\geq t)+\expit(\lambda_n(-\delta/2))\P(\eta_a(X)<t)\to \P(\eta_a(X)\geq t)\quad (n\to \infty)
\end{align*}
This implies $\mu_a <\mu_a$, which is a contradiction. Thus,
\[
\limsup_{\lambda_a\to\infty}c_{\lambda_a}\leq c_\mathrm{max}.
\]

From \[
\liminf_{\lambda_a\to\infty}c_{\lambda_a}\geq c_\mathrm{min}\text{ and }\limsup_{\lambda_a\to\infty}c_{\lambda_a}\leq c_\mathrm{max}
\]
we can conclude 
\[
p_{a,\lambda_a}(X)=\expit(\lambda_a(\eta_a(X)-c_{\lambda_a}))\to1 \qquad \text{on } \{\eta_a(X)>c_\mathrm{max}\}
\]
and
\[
p_{a,\lambda_a}(X)=\expit(\lambda_a(\eta_a(X)-c_{\lambda_a}))\to0 \qquad \text{on } \{\eta_a(X)<c_\mathrm{min}\}.
\]
Furthermore, by definition of $\mathcal C$, \[
\P(c_\mathrm{min}<\eta_a(X)<c_\mathrm{max})=\P(c_\mathrm{min} <\eta_a(X))-\P(\eta_a(X)\geq c_\mathrm{max})\leq \mu_a-\mu_a=0,\] 
so $\{c_\mathrm{min}<\eta_a(X)<c_\mathrm{max}\}$ is a $\P$-nullset.

To complete the proof of the first part of Proposition 3, it remains to determine the behaviour of $p_{a,\lambda_a}$ on $\{\eta_a(X)=c_\mathrm{min}\}$ and $\{\eta_a(X)=c_\mathrm{max}\}$.
We distinguish two cases.

Case 1): Suppose $c_\mathrm{\min}<c_\mathrm{max}$. Then 
\[
\P(\eta_a(X)>c_\mathrm{min})\leq \mu_a\leq \P(\eta_a(X)\geq c_\mathrm{max})\leq \P(\eta_a(X)>c_\mathrm{min}),
\]
where the last inequality follows from the assumption. Consequently, 
\[
\P(\eta_a(X)>c_\mathrm{min})=\mu_a=\P(\eta_a(X)\geq c_\mathrm{max}).
\]
Further,
\begin{align*}
\mu_a
&=\E\{\expit(\lambda_a(\eta_a(X)-c_{\lambda_a}))\}\\
&=\begin{aligned}[t]&\E\{\expit(\lambda_a(\eta_a(X)-c_{\lambda_a}))\mathbf 1_{\{\eta_a(X)<c_\mathrm{min}\}}\}\\
&+\E\{\expit(\lambda_a(\eta_a(X)-c_{\lambda_a}))\mathbf 1_{\{\eta_a(X)=c_\mathrm{min}\}}\}\\
&+\E\{\expit(\lambda_a(\eta_a(X)-c_{\lambda_a}))\mathbf 1_{\{c_\mathrm{min}<\eta_a(X)<c_\mathrm{max}\}}\}\\
&+\E\{\expit(\lambda_a(\eta_a(X)-c_{\lambda_a}))\mathbf 1_{\{\eta_a(X)=c_\mathrm{max}\}}\}\\
&+\E\{\expit(\lambda_a(\eta_a(X)-c_{\lambda_a}))\mathbf 1_{\{\eta_a(X)>c_\mathrm{max}\}}\}.\end{aligned}
\end{align*}
As the $\expit$ function is bounded by 0 and 1, dominated convergence and our previous results imply
\begin{align*}
\mu_a
&=\begin{aligned}[t]&\P\{\eta_a(X)>c_\mathrm{max}\}\\
&+\lim_{\lambda_a\to\infty}\P\{\eta_a(X)=c_\mathrm{min}\}\expit(\lambda_a(c_\mathrm{min}-c_{\lambda_a}))+\P\{\eta_a(X)=c_\mathrm{max}\}\expit(\lambda_a(c_\mathrm{max}-c_{\lambda_a}))\end{aligned}\\
&=\mu_a+\lim_{\lambda_a\to\infty}\P\{\eta_a(X)=c_\mathrm{min}\}\expit(\lambda_a(c_\mathrm{min}-c_{\lambda_a}))-\P\{\eta_a(X)=c_\mathrm{max}\}\{1-\expit(\lambda_a(c_\mathrm{max}-c_{\lambda_a}))\},
\end{align*}
where in the last step we used $\P(\eta_a(X)\geq c_\mathrm{max})=\mu_a$. It follows that 
\[
\lim_{\lambda_a\to\infty}\P\{\eta_a(X)=c_\mathrm{min}\}\expit(\lambda_a(c_\mathrm{min}-c_{\lambda_a}))-\P\{\eta_a(X)=c_\mathrm{max}\}\{1-\expit(\lambda_a(c_\mathrm{max}-c_{\lambda_a}))\}=0.
\]
Moreover, elementary algebraic manipulations yield 
\begin{align*}
    &\expit(\lambda_a(c_\mathrm{min}-c_{\lambda_a}))\{1-\expit(\lambda_a(c_\mathrm{max}-c_{\lambda_a}))\}\\&=e^{-\lambda_a(c_\mathrm{max}-c_\mathrm{min})}\{1-\expit(\lambda_a(c_\mathrm{min}-c_{\lambda_a}))\}\expit(\lambda_a(c_\mathrm{max}-c_{\lambda_a})),
\end{align*}
and thus \[
\expit(\lambda_a(c_\mathrm{min}-c_{\lambda_a}))\{1-\expit(\lambda_a(c_\mathrm{max}-c_{\lambda_a}))\}\leq e^{-\lambda_a(c_\mathrm{max}-c_\mathrm{min})}\to 0\quad (\lambda_a\to\infty),
\]
implying
\[
\lim_{\lambda_a\to\infty}\P\{\eta_a(X)=c_\mathrm{min}\}\expit(\lambda_a(c_\mathrm{min}-c_{\lambda_a}))\cdot\P\{\eta_a(X)=c_\mathrm{max}\}\{1-\expit(\lambda_a(c_\mathrm{max}-c_{\lambda_a}))\}= 0.
\]
Therefore, in the last expression, the left or the right product has to converge to 0. However, as previously shown, also the difference of the left and right product goes to 0, and thus both have to go to 0 as $\lambda_a\to\infty$. Hence, if $\P(\eta_a(X)=c_\mathrm{min})>0$, then 
\[
 \expit(\lambda_a(c_\mathrm{min}-c_{\lambda_a}))\to0\quad (\lambda_a\to\infty) \
\]
and if $\P(\eta_a(X)=c_\mathrm{max})>0$, then 
\[
\expit(\lambda_a(c_\mathrm{max}-c_{\lambda_a}))\to1\quad (\lambda_a\to\infty).
\]

In conclusion, 
\[
p_{a,\lambda_a}(X) \to\mathbf 1_{\{\eta_a(X)> c_\mathrm{max}\}}+\mathbf{1}_{\{\eta_a(X)=c_\mathrm{max}\}} \quad (\lambda_a\to\infty)\text{ almost surely},
\]
which, because of $\P(c_\mathrm{min}<\eta_a(X)<c_\mathrm{max})=0$, $\expit(\lambda_a(c_\mathrm{min}-c_{\lambda_a}))\xrightarrow{\lambda_a\to\infty}0$ and $\P(\eta_a(X)\geq c_\mathrm{max})=\mu_a$, implies 
\[
p_{a,\lambda_a}(X)\to \mathbf 1_{\{\eta_a(X)>c\}}+q1_{\{\eta_a(X)=c\}}\quad (\lambda_a\to \infty)\text{ almost surely}
\] for every $c\in \mathcal C$, where $q$ is defined as in the statement of Proposition 3.

Case 2): Suppose $c_\mathrm{min}=c_\mathrm{max}$. Let $c\in \mathcal C$. Then 
\begin{align*}
\mu_a&=\begin{aligned}[t]&\E\{\expit(\lambda_a(\eta_a(X)-c_{\lambda_a}))\mathbf 1_{\{\eta_a(X)>c\}}\}\\
&+\E\{\expit(\lambda_a(\eta_a(X)-c_{\lambda_a}))\mathbf 1_{\{\eta_a(X)=c\}}\}\\
&+\E\{\expit(\lambda_a(\eta_a(X)-c_{\lambda_a}))\mathbf 1_{\{\eta_a(X)<c\}}\}.\\
\end{aligned}
\end{align*}
Taking the limit yields 
\[
\mu_a=\P(\eta_a(X)>c)+\P(\eta_a(X)=c)\cdot \lim_{\lambda_a\to \infty}\expit(\lambda_a(\eta_a(X)-c_{\lambda_a}))+0,
\]
i.e., if $\P(\eta_a(X)=c)>0$,
\[
\expit(\lambda_a(c-c_{\lambda_a}))\to\frac{\mu_a-\P(\eta_a(X)>c)}{\P(\eta_a(X)=c)}\quad (\lambda_a\to\infty).
\]
Consequently, \[
p_{a,\lambda_a}(X)\to \mathbf 1_{\{\eta_a(X)>c\}}+q1_{\{\eta_a(X)=c\}}\quad (\lambda_a\to \infty)\text{ almost surely},
\]
which concludes the proof of the first part of Proposition 3.
The second part now follows immediately by $\Var(p_{a,\lambda}(X))=\E(p_{a,\lambda_a}(X)^2)-\E(p_{a,\lambda_a}(X))^2$ and dominated convergence.
\end{proof}

\subsection{Why Marginal Prognostic Strength is the Calibration Coordinate}

The calibration procedure uses control-arm marginal prognostic strength to index the structure-preserving path and treats precision-gain potential as an estimand-specific output. 
Since precision-gain potential is ultimately the quantity of interest when evaluating the efficiency benefit of covariate adjustment, a natural question is whether it could instead be used directly as the calibration coordinate. 

Like marginal prognostic strength, precision-gain potential has a well-defined limiting value along the structure-preserving path. For the remainder of this subsection, let $\mathcal G^{\mathrm{abs},\lambda}_{\pi,\psi}(X)$ and $\mathcal G^{\mathrm{rel},\lambda}_{\pi,\psi}(X)$ denote, respectively, the absolute and relative precision-gain potentials under the data-generating mechanism $(p_{0,\lambda},p_{1,\lambda})\in\mathcal P$.

\begin{coro}[Limiting precision-gain potential] 
\label{coro:pgp-limit} 
Let $\psi=f(\mu_1,\mu_0)$ satisfy the assumptions of Proposition~1. 
For each $a\in\{0,1\}$, suppose that the assumptions of Proposition~3 hold for the arm-specific score $\eta_a(X)$ and marginal risk $\mu_a$. 
Let $c_a$ and $q_a$ denote, respectively, the quantities $c$ and $q$ from Proposition~3 when that proposition is applied to treatment arm $a$, and define
\[ 
p_{a,\infty}(X) = \mathbf 1_{\{\eta_a(X)>c_a\}} + q_a\mathbf 1_{\{\eta_a(X)=c_a\}}, \qquad a\in\{0,1\}. 
\] 
Then
\[ 
\lim_{\lambda\to\infty} \mathcal G^{\mathrm{abs},\lambda}_{\pi,\psi}(X) = \Var\left( \sqrt{\frac{1-\pi}{\pi}}\,f_1p_{1,\infty}(X) - \sqrt{\frac{\pi}{1-\pi}}\,f_0p_{0,\infty}(X) \right), 
\]
and 
\[ 
\lim_{\lambda\to\infty} \mathcal G^{\mathrm{rel},\lambda}_{\pi,\psi}(X) = \frac{ \Var\left( \sqrt{\frac{1-\pi}{\pi}}\,f_1p_{1,\infty}(X) - \sqrt{\frac{\pi}{1-\pi}}\,f_0p_{0,\infty}(X) \right) }{ \nu_{\mathrm{unadj}}(\psi) }. 
\] 
\end{coro}
\begin{proof} 
For each $a\in\{0,1\}$, Proposition 3, applied to the arm-specific score $\eta_a(X)$ and marginal risk $\mu_a$, yields 
\[ p_{a,\lambda}(X) \longrightarrow p_{a,\infty}(X) \qquad\text{almost surely}. \]
Consequently, the $\lambda$-indexed random variable 
\[ \sqrt{\frac{1-\pi}{\pi}}\,f_1p_{1,\lambda}(X) - \sqrt{\frac{\pi}{1-\pi}}\,f_0p_{0,\lambda}(X) \] 
converges almost surely to 
\[ \sqrt{\frac{1-\pi}{\pi}}\,f_1p_{1,\infty}(X) - \sqrt{\frac{\pi}{1-\pi}}\,f_0p_{0,\infty}(X). \] 
Because $0\leq p_{a,\lambda}(X)\leq1$, the absolute value of the $\lambda$-indexed random variable is uniformly bounded by \[ \sqrt{\frac{1-\pi}{\pi}}\,|f_1| + \sqrt{\frac{\pi}{1-\pi}}\,|f_0|. \] 
Dominated convergence therefore implies convergence of its first and second moments and hence of its variance.
By Proposition 1, this variance is $\mathcal G^{\mathrm{abs},\lambda}_{\pi,\psi}(X)$, which proves the asserted limit for absolute precision-gain potential. 
The arm-specific marginal risks remain fixed along the structure-preserving path.
Consequently, $\nu_{\mathrm{unadj}}(\psi)$ does not depend on $\lambda$. 
Dividing the absolute precision-gain limit by this constant proves the result for relative precision-gain potential. 
\end{proof} 

Corollary~\ref{coro:pgp-limit} shows that precision-gain potential has a well-defined endpoint along the structure-preserving path. 
A unique calibration coordinate, however, requires more than the existence of endpoint values: it must vary monotonically along the path. Precision-gain potential does not generally satisfy this requirement.

To demonstrate this distinction, we apply Algorithm~1 to a data-generating mechanism for which precision-gain potential is nonmonotone along the resulting path. 
Let
\[X\sim \bernoulli(0.6),\quad \quad \mu_0=0.6, \quad \mu_1=0.3, \quad \pi=\frac{1}{2},\] 
and 
\[h(X)=2X-0.9,\quad v_\mathrm{hom}=-1,\quad v_\mathrm{het}=5. \]

\begin{figure}[h]
    \centering
   \includegraphics[width=1.0\linewidth]{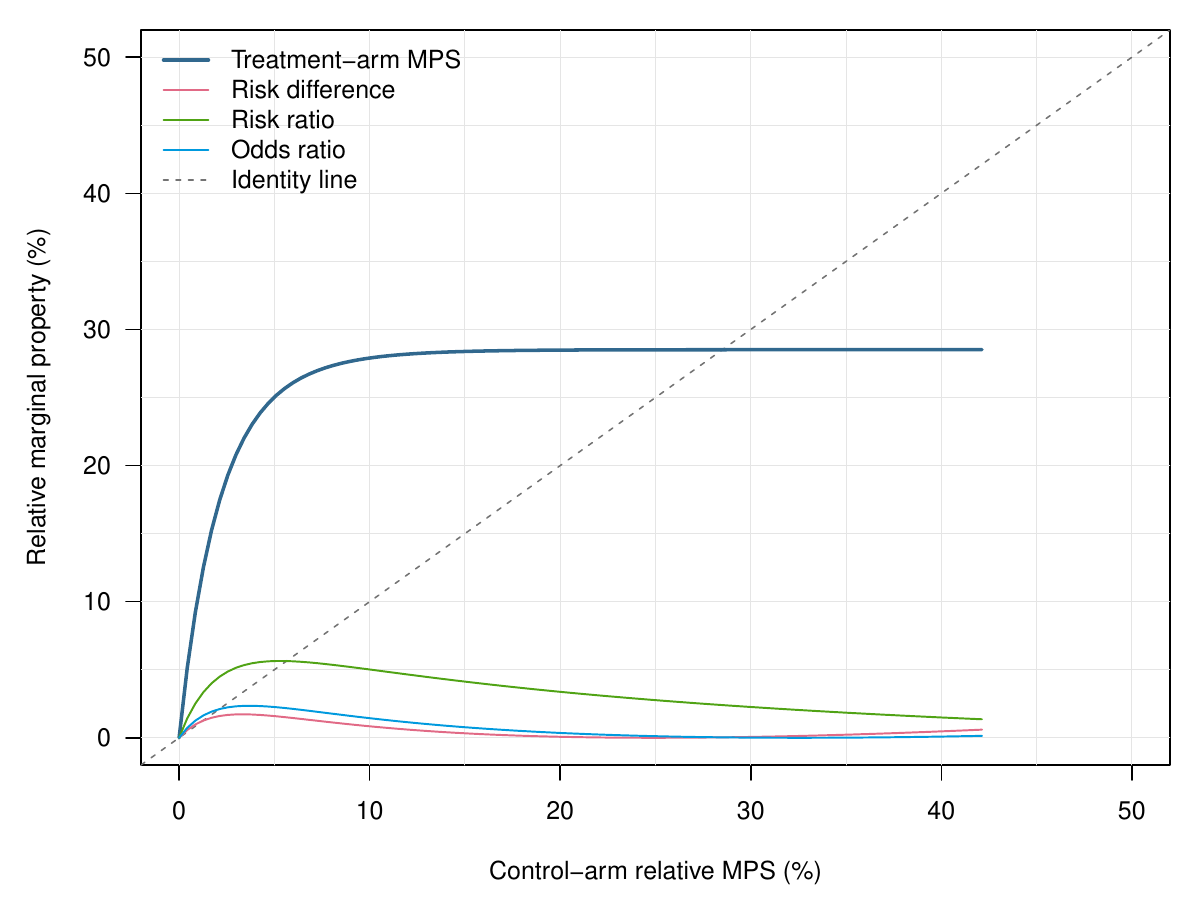} 
    \caption{Structure-preserving calibration path illustrating that precision-gain potential need not be monotone. Although marginal prognostic strength increases monotonically in both treatment arms, the relative precision-gain potentials are nonmonotone along the path.}
    \label{fig:pgp-nonmonotone}
\end{figure}

Figure~\ref{fig:pgp-nonmonotone} shows that, unlike control-arm marginal prognostic strength, $\mathcal G^{\mathrm{rel},\lambda}_{\pi,\psi}(X)$ need not be monotone along the structure-preserving path. Consequently, a given precision-gain target may correspond to multiple common multipliers and hence to multiple data-generating mechanisms. Precision-gain potential therefore does not generally provide a unique calibration coordinate.

\end{document}